\documentclass[sigconf]{aamas} 
\usepackage{amsmath}
\usepackage{bm}

\usepackage{balance} 

\setcopyright{ifaamas}
\acmConference[AAMAS '23]{Proc.\@ of the 22nd International Conference
on Autonomous Agents and Multiagent Systems (AAMAS 2023)}{May 29 -- June 2, 2023}
{London, United Kingdom}{A.~Ricci, W.~Yeoh, N.~Agmon, B.~An (eds.)}
\copyrightyear{2023}
\acmYear{2023}
\acmDOI{}
\acmPrice{}
\acmISBN{}

\title[]{Artificial prediction markets present a novel opportunity for human-AI collaboration }

\author{Tatiana Chakravorti}
\affiliation{
  \institution{Pennsylvania State University}
  \city{State College}
  \country{USA}}
\email{tfc5416@psu.edu}

\author{Vaibhav Singh}
\affiliation{
  \institution{Pennsylvania State University}
  \city{State College}
  \country{USA}}
\email{vxs5308@psu.edu}

\author{Sarah Rajtmajer}
\affiliation{
  \institution{Pennsylvania State University}
  \city{State College}
  \country{USA}}
\email{smr48@psu.edu}

\author{Michael McLaughlin}
\affiliation{
  \institution{Pennsylvania State University}
  \city{State College}
  \country{USA}}
\email{mvm7085@psu.edu}

\author{Robert Fraleigh}
\affiliation{
  \institution{Pennsylvania State University}
  \city{State College}
  \country{USA}}
\email{rdf5090@psu.edu}

\author{Christopher Griffin}
\affiliation{
  \institution{Pennsylvania State University}
  \city{State College}
  \country{USA}}
\email{cxg286@psu.edu}

\author{Anthony Kwasnica}
\affiliation{
  \institution{Pennsylvania State University}
  \city{State College}
  \country{USA}}
\email{amk17@psu.edu}

\author{David Pennock}
\affiliation{
  \institution{Rutgers University}
  \city{New Jersey}
  \country{USA}}
\email{david.pennock@rutgers.edu}

\author{C. Lee Giles}
\affiliation{
  \institution{Pennsylvania State University}
  \city{State College}
  \country{USA}}
\email{clg20@psu.edu}

\begin{abstract}
Despite high-profile successes in the field of Artificial Intelligence, machine-driven technologies still suffer important limitations, particularly for complex tasks where creativity, planning, common sense, intuition, or learning from limited data is required. These limitations motivate effective methods for human-machine collaboration. Our work makes two primary contributions. We thoroughly experiment with an artificial prediction market model to understand the effects of market parameters on model performance for benchmark classification tasks. We then demonstrate, through simulation, the impact of exogenous agents in the market, where these exogenous agents represent primitive human behaviors. This work lays the foundation for a novel set of hybrid human-AI machine learning algorithms.
\end{abstract}

\keywords{prediction markets; machine learning; artificial intelligence; human-AI collaboration}

\newcommand{\BibTeX}{\rm B\kern-.05em{\sc i\kern-.025em b}\kern-.08em\TeX}

\begin{document}


\pagestyle{fancy}
\fancyhead{}


\maketitle


\section{Introduction}

A body of work on artificial prediction markets is emerging. These are numerically simulated markets, populated by artificial agents (bot-traders) for the purpose of supervised learning of probability estimators \cite{barbu2012introduction}. While nascent, this literature has demonstrated the plausibility of using a trained market as a supervised learning algorithm, achieving comparable performance to standard approaches on simple classification tasks \cite{barbu2012introduction,barbu2013artificial,jahedpari2014artificial,nakshatri2021design}. In fact, these results are sensible given the deep mathematical connections between prediction markets and learning  \cite{chen2008complexity,chen2010new,abernethy2011optimization}. 

Like other machine learning algorithms, functioning of an artificial prediction market depends on several researcher-determined parameters: number of agents; liquidity; initial cash; alongside parameters related to training processes. Scenarios in which performance is robust or brittle to these settings is yet unclear. Prior work has observed that artificial markets may suffer from lack of participation \cite{rajtmajer2022synthetic}. That is, like their human counterparts in traditional prediction markets, agents may not invest in the market if they do not have sufficient information \cite{Arrow:2008,tetlock2008liquidity,rothschild2014extent}; in practice, this occurs when an asset representing test data point is too dissimilar to training examples.

In our view, the most promising opportunity afforded by artificial prediction markets is eventual human-AI collaboration -- a market framework should theoretically support human traders participating alongside agents to evaluate outcomes. Whether and how artificial prediction markets might benefit from this hybrid scenario is an open question. The work we undertake here provides, through simulation, initial support for this opportunity in the context of a simple artificial market and primitive human behaviors.

\noindent Our work is framed by two primary research questions.

\textbf{RQ1}: How does performance of a simple artificial prediction market depend on hyper-parameter selection?

\textbf{RQ2}: What impact does the inclusion of exogenous agents representing simple (human-like) behaviors have on market performance?

Our findings support those of prior recent work indicating the promise of artificial prediction markets for classification tasks. We demonstrate the sensitivity of this approach to hyper-parameter selection and highlight, in particular, the role of liquidity in moderating performance. Finally, we demonstrate the exciting opportunity for \emph{hybrid prediction markets} to serve as a framework for human-AI collaboration. We suggest that this approach may be particularly valuable in contexts where machine learning falls short (e.g., lack of training data, complex tasks) and potential for human-only approaches is either undesirable or infeasible.


\section{Related Work}

Our work builds upon and contributes to two primary literatures, namely, work on artificial prediction markets and work in collaborative human-AI technologies.

\subsection{Artificial Prediction Markets}

Prediction markets are simple futures markets used to aggregate disperse information into efficient forecasts of uncertain future events \cite{wolfers2004prediction,hanson2006information,manski2006interpreting, wolfers2006interpreting}. Specifically, market participants buy and sell contracts that pay out based on the outcomes of future events. Market prices generated from these contracts can be understood as a collective prediction among market participants. 
Prediction markets have been successfully used, e.g., for forecasting election outcomes \cite{berg2008prediction}, sports betting \cite{spann2009sports}, forecasting infectious disease activity \cite{polgreen2007use}, and aggregating employee wisdom in corporate settings \cite{cowgill2009using,gillen2012information}. 

Artificial prediction markets are a variation on this idea, wherein numerically simulated markets populated by trained agents (bot-traders) are used for the purpose of supervised learning of probability estimators \cite{barbu2012introduction, barbu2013artificial}.
In initial formulations by Barbu and Lay \cite{lay2010supervised,barbu2012introduction,lay2012artificial}, each agent is represented as a budget and a simple betting function. During training, each agent's budget is updated based on the accuracies of its predictions over a training dataset. 
Authors found that these markets outperformed standard approaches on benchmark classification and regression tasks. 
Later, Storkey and colleagues \cite{storkey2011machine,storkey2012isoelastic} developed 
the so-called machine learning market, also for the purpose of classification. In their formulation, each agent purchases contracts in order to maximize a utility function. 
Most recently, Nakshatri et al. \cite{nakshatri2021design} proposed an artificial prediction market wherein agent purchase logic is defined geometrically, in particular, by a convex semi-algebraic set in feature space. Time varying asset prices affect the structure of the semi-algebraic sets leading to time-varying agent purchase rules. Agent parameters are trained using an evolutionary algorithm. Authors show that their approach has desirable properties, e.g., the market satisfies certain universal approximation properties, and there exist sufficient conditions for convergence. Our work builds on this approach.

Like their human-populated counterparts, artificial prediction markets have found a number of real-world applications \cite{barbu2013artificial,jahedpari2014artificial}. Ongoing theoretical work has offered support for these promising experimental findings, highlighting the mathematical connections between artificial markets and machine learning \cite{chen2008complexity,ChenPennock:2010,abernethy2011optimization, hu2014multi}.

\subsection{Human-AI Collaboration}

Despite high-profile successes in the field of Artificial intelligence (AI) \cite{he2015delving,brown2019superhuman,kleinberg2018human,zhu2018human}, machine-driven solutions still suffer important limitations particularly for complex tasks where creativity, common sense, intuition or learning from limited data is required \cite{jarrahi2018artificial,lai2019human,green2019principles,li2016crowdsourced,kamar2016directions,amershi2019guidelines,muller2017organic}. Both the promises and challenges of AI have motivated work on human-machine collaboration \cite{dellermann2021future,wang2019human,nunes2015survey,puig2020watch,wu2022survey}. The hope is that we can eventually develop hybrid systems that bring together human intuition and machine rationality to effectively and efficiently tackle today’s grand challenges. 

Recent work in hybrid intelligence systems has demonstrated the feasibility and highlighted the potential of integrating human input into AI systems \cite{kamar2016directions}, or even, of human-AI collaboration \cite{wang2020human}. The spectrum of these efforts range from accounting for human factors in technology design \cite{bansal2019beyond,canonico2019wisdom,harper2019role} to efficiently utilizing human inputs for training data \cite{amershi2014power} in applications as diverse as business \cite{nagar2011making,sowa2021cobots}, civic welfare \cite{fogliato2022case}, criminal justice \cite{travaini2022machine}, and healthcare \cite{tschandl2020human,lee2021human,rajpurkar2022ai}. 

The work we describe here brings together the bodies of prior work on artificial prediction markets and hybrid intelligence, proposing hybrid prediction markets for direct integration of human wisdom into the deployment of a machine learning algorithm.

\section{Data}

We consider three classification tasks.  The first two are benchmark tasks used broadly to compare performance of machine learning algorithms.  The third is the task of classifying scientific research outcomes as replicable or not replicable -- a challenging, complex task on which both machine learning algorithms \cite{altmejd2019predicting,yang2020estimating,pawel2020probabilistic,wu2021predicting} and human assessment \cite{dreber2015using,camerer2016evaluating,camerer2018evaluating,forsell2019predicting,gordon2020replication,gordon2021predicting} have achieved respectable but not excellent performance. The replication prediction task, we suggest, is an example of the type of problem well-suited to hybrid human-AI approaches.  

\subsection{Benchmark Machine Learning Datasets}
\label{benchmark_datasets}

The Iris dataset \cite{misc_iris_53} was one the earliest datasets used for evaluation of classification methodologies. The dataset contains three classes of $50$ instances each, where each class refers to a type of iris plant. One class is linearly separable from the others; the latter are not linearly separable from one another. For evaluation using the binary market, we have combined the latter two classes (iris virginica and iris versicolor). Prior approaches for classification of the Iris dataset based on support vector classification \cite{mohan2020support}, random forest classification \cite{mishina2015boosted, chicho2021machine}, and logisitc regression \cite{pinto2018iris} have reported $100\%$ or near-$100\%$ accuracy on the task.  

In addition to the Iris dataset, we consider the Heart Disease dataset \cite{misc_heart_disease_45}. The Heart Disease dataset is also a multivariate dataset used for benchmark classification algorithms. Fourteen patient attributes are used to predict presence or absence of heart disease. Random forest \cite{singh2016heart}, Xgboost \cite{rajadevi2021feature}, and logistic regression \cite{desai2019back} achieve performance just under $90\%$ accuracy. While, support vector classification achieves $86\%$ \cite{rajadevi2021feature}.

\subsection{Replication Studies Outcomes}

In the last decade, several large-scale replication projects have been undertaken across psychology, economics, political science, cancer biology and other domains \cite{open2015estimating,camerer2016evaluating,camerer2018evaluating,klein2014investigating,klein2018many,cova2021estimating,errington2014open}. Amongst their important impacts, these studies have created small ground-truth datasets of replication studies outcomes that can be used for train and test of automated approaches for replication prediction. Specifically, we use the dataset and extracted features considered by \cite{rajtmajer2022synthetic} for ease of comparison. The dataset containes 192 findings in the social and behavioral sciences, each labeled either Replicable or Not Replicable, and a set of 41 features extracted from each associated paper representing biblometric, venue-related, author-related, statistical and semantic information. See \cite{wu2021predicting} for further detail on feature extraction processes.

Of note, authors in \cite{rajtmajer2022synthetic} achieve $89.4\%$ accuracy, remarkable for the task of replication prediction. However, accuracy is calculated based on the approximately one-third of the test data that gets evaluated by the market. Because agent participation is voluntary and agents do not participate if they do not have sufficient information about a test point, some (or much) of the data can be left unclassified. Our work uses the same data and market structure described in \cite{rajtmajer2022synthetic}. This allows us to explore the effects of hyper-parameters (RQ1) and the inclusion of exogenous agents (RQ2) on these performance/participation trade-offs.

\section{Prediction Market Model}

We use as a base model the artificial binary prediction market described in \cite{nakshatri2021design}. The state of the prediction market is defined by a pair of integers $\mathbf{q}_t = (q^0_t,q^1_t) \in \mathbb{Z}_+^2$ giving the number of units of the two asset classes that have been sold. For simplicity we refer to the assets as $0$ and $1$. Traders are agents $\mathcal{A} = \{a_1,\dots,a_n\}$ who buy assets $0$ and $1$ using policies $\{\gamma_1,\dots,\gamma_n\}$. Also following \cite{nakshatri2021design}, we assume for simplicity that agents cannot sell. 
If agent purchase policy $\gamma_i$ is conditioned on exogenous information $\mathbf{x} \in D \subseteq \mathbb{R}^n$ then, $\gamma_i : (\mathbf{q}_t,\mathbf{x}) \mapsto (r^0,r^1)$ and agent $i$ purchases $r^0$ units of $A_0$ and $r^1$ units of $A_1$, thus causing a state update. In what follows, we assume that agents specialize in the purchase of either Asset $0$ or Asset $1$ so that if $r^0 > 0$, then $r^1 = 0$. 

Asset prices are computed using a logarithmic market scoring rule (LMSR): 
\begin{gather*}
p^0_t = \frac{\exp{(\beta q^0_t)}}{\exp{(\beta q^0_t)} + \exp{(\beta q^1_t)}} \\
p^1_t = \frac{\exp{(\beta q^1_t)}}{\exp{(\beta q^0_t)} + \exp{(\beta q^1_t)}}.
\end{gather*}
This is the softmax function 
of $(q^0_t,q^1_t)$. Liquidity $\beta$ adjusts the price change given a change in asset quantities \cite{lekwijit2018optimizing}. The fact that prices vary as a function of $\mathbf{q}_t$ ensures that the policy need not take spot price into consideration explicitly. It is often more convenient to work in units of $1/\beta$ as $\beta$ can become arbitrarily close to zero. Experimental results are therefore reported for this quantity as the \textit{liquidity factor}. 

To start the market, all agents may purchase assets at time $t = 0$. After this, we assume that agents arrive at the market with arrival rate $\lambda$ and inter-arrival time governed by an exponential distribution. This allows us to avoid scenarios in the hybrid setting where the synthetic traders swamp the market.

The LMSR imposes a market maker price, so that actual trade costs are given by:
\begin{gather*}
\kappa_t^0(\Delta q^0) = \frac{1}{\beta} \log\left\{\frac{\exp[\beta (q^0_t+\Delta q^0)] + \exp[\beta q^1_t]}{\exp[\beta q^0_t] + \exp[\beta q^1_t]}\right\}\\
\kappa_t^1(\Delta q^1) = \frac{1}{\beta} \log\left\{\frac{\exp[\beta q^0_t] + \exp[\beta (q^1_t + \Delta q^1)]}{\exp[\beta q^1_t] + \exp[\beta q^0_t]}\right\}.
\end{gather*}
Here $\kappa^i_t(\Delta q^i)$ is the cost to a trader for purchasing $\Delta q^i$ units of Asset $i$ (with $i \in \{0,1\})$ at time $t$. For small values of $\beta$ (large values of $1/\beta$) the cost of purchase approaches the spot-price \cite{nakshatri2021design}.

Agent purchase logic is governed by a time-varying bank value $B_i$ and a characteristic function $\psi_i:\mathbb{R}^n \times \mathbb{R}^2\times\mathbb{R}^m \to \mathbb{R}$ to reason about information $\mathbf{x}$ and its decision to buy an asset in class $y_i$ is governed by:
\begin{equation}
\Delta q^{y_i}_i = H\left\{\sigma[\psi_i(\mathbf{x},\mathbf{q};\bm{\theta})] - \kappa^{y_i}\right\}\cdot H\left(B_i - \kappa^{y_i}\right).
\label{eqn:PurchaseEqn}
\end{equation}
Here $\sigma:\mathbb{R}\to[0,1]$ is a sigmoid function and $H(x)$ is the unit step function defined as $0$ at $x = 0$. The expression $\sigma[\psi_i(\mathbf{x},\mathbf{q})]$ defines the value Agent $i$ places on Asset $y_i$ as a function of the market state (and hence spot-prices) and the information in the external information $\mathbf{x}$. If Agent $i$ places more value on Asset $y_i$ than its present price $\kappa^{y_i}$, then $H\left\{\sigma[\psi_i(\mathbf{x},\mathbf{p})] - \kappa^{y_i}\right\} = 1$ and $\Delta q^{y_i}_i = 1$ just in case the agent has sufficient funds given by $H\left[B_i - \kappa^{y_i}\right]$. That is, Agent $i$ purchases a share of Asset $i$. Notice we are assume that agents may buy one share of an asset at a time. This both simplifies the agent logic and also would prevent the agents from out-competing humans in the market in the hybrid scenario. The vector $\bm{\theta}$ is a set of parameters that define the specific outputs of $\psi_i$ and thus affect the agent purchase logic.

Let $\bm{\Theta}$ be the (matrix) of all parameter vectors for the agents. After running for $T$ time units with input information $\mathbf{x}$, the spot price for Asset $1$ is $p^1_T(\mathbf{x};\bm{\Theta})$. If we are given input information $\{\mathbf{x}_1,\dots,\mathbf{x}_N\}$ with class information $\{y_1,\dots,y_N\}$, then training the market is the process of solving:
\begin{equation*}
\min_{\bm{\Theta}}\;\;\frac{1}{N}\sum_{j=1}^N \left\lvert p^1_T(\mathbf{x}_j;\bm{\Theta}) - y_j\right\rVert^2.
\end{equation*}
This problem is solved in \cite{nakshatri2021design} using a genetic algorithm to obtain a market that can classify external information $\mathbf{x} \in D$. 

At the close of the market, the price of a each asset is taken as a proxy for the market's confidence in the corresponding outcome. In our binary market model, there are two mutually exclusive possible outcomes and so the (normalized) prices should sum to $1$. In this way, the market can be used for regression or classification. In the three examples we consider here, the market is used for classification. A separate market is run for each point in the test set and the asset with the higher price is considered the market's classification decision for that test point.

We note, critically, that based on this model, agent participation is voluntary and decision to participate is driven by $\Delta q^{y_i}_i = 1$ from Equation \ref{eqn:PurchaseEqn}. If this condition is not met during the course of the market for any agents, there will be no market activity and thus no classification decision for that test point. Authors in \cite{rajtmajer2022synthetic} have noted that this may occur frequently, particularly in cases where the training data set is small or points in the test set are significant different from training the data. Accordingly, we calculate accuracy and F1 based on the scored subset of the data, while also reporting the percentage of scored test data as a performance metric.  

The artificial prediction market includes five hyper-parameters that are not optimized by the genetic algorithm discussed in \cite{nakshatri2021design}:
\begin{enumerate}
\item Agent inter-arrival rate ($\lambda$);
\item Agent initial bank value ($B_i(0)$);
\item Market liquidity ($1/\beta$);
\item Simulation running time ($T$) or duration;
\item Number of generations in the genetic (training) algorithm.
\end{enumerate}

\noindent As such, these parameters are researcher-determined and warrant further study (RQ1). Our first set of experiments, described below, explore the specific roles of agent inter-arrival rate ($\lambda$), agent initial bank value ($B_i(0)$) (or, ``cash''), and market liquidity ($1/\beta$) on performance. We explore the robustness of performance to selection of these hyper-parameters, highlighting accuracy and F1 score but also trade-offs with agent participation.

In experiments that follow, the genetic algorithm is trained over five generations. The objective function of the genetic algorithm maximizes root mean square error of the estimated score. Agent performance is evaluated based on profit; nonprofitable agents are deleted from the pool. The ten most profitable agents are retained and, amongst them, the seven most profitable agents are selected for mutation and crossover.

\section{Experimental Design}

The following experiments support the two primary research questions we have put forward. First, we capture the effects of different combinations of hyper-parameters on market performance (RQ1). Second, we explore the impact of exogenous agents not trained through the evolutionary training process, but rather who adopt one of a set of three simple purchasing rules meant to represent primitive human inputs (RQ2).

\subsection{Market robustness to hyper-parameters}

\begin{figure}[h]
  \centering
  \includegraphics[width=0.95\linewidth]{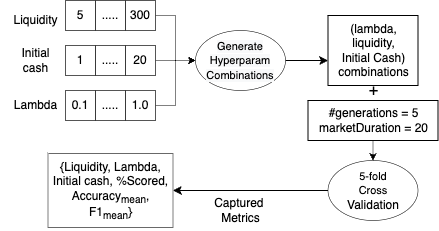}
  \caption{RQ1 experimental architecture.}
  \label{phase_1_arch}
\end{figure}

We study the effects of inter-arrival rate $\lambda$, agent initial bank value $B_i(0)$ (or, ``cash''), and market liquidity factor $1/\beta$ on artificial market performance.  As mentioned, number of generations is fixed at five during training; while, market duration is fixed at 20. These parameters were fixed (vs. manipulated) to avoid combinatorial complexity during this initial study; however, they should be further studied in future work. In practice, we have found these values to be sufficient for market behavior to converge while also offering reasonable run time. 
Liquidity factor is tested for the set of values $\{5, 10, 20, 50, 75, 100, 150, 200, 300\}$. Initial cash is tested for $\{1, 2, 3, 4, 5, 10, 20\}$; $\lambda$ is tested for $\{0.01, 0.025, 0.05, 0.1, 0.25, 0.5, 1.0\}$. Our experiments consider all combinations of these hyper-parameter values, $441$ total, and measure corresponding performance in terms of accuracy, F1 score, and percentage of scored test points. Performance for each hyper-parameter set is determined based on 5-fold cross validation with 80/20 train/test splits and performance metrics are averaged over the folds. From these outcomes, we select best and worst-performing hyper-parameter sets to be used for downstream analyses.
This process is outlined in Figure \ref{phase_1_arch}.

\vspace{-0.2cm}

\subsection{Market behavior with exogenous agents}

We introduce three classes of exogenous agents representing simple, fundamental behaviors which operate fully separate from the agent logic and feature-based training protocol used for the other agents in the market.  These classes of behavior are intended to represent behavioral primitives that, in combination, would underlie the actions of human participants in a hybrid scenario.
The first, \emph{ground truth} agents (GT) have perfect knowledge of the correct outcome and always buy contracts corresponding to the correct outcome whenever they have the opportunity to participate (which is moderated by their arrival rate, $\lambda$).
The second are \emph{ground truth inverse} agents (GTinv). These agents also know the correct outcome but always buy contracts corresponding to the incorrect outcome whenever they have an opportunity to participate. This scenario is equivalent to the case where agents are simply certain but incorrect in their forecast. Finally, our third class of agents are \emph{random} agents which purchase contracts corresponding to one or the other outcome randomly. Understanding that the decisions of human participants in the hybrid prediction market would not fall squarely into these three categories, these simulations are intended to draw initial boundaries around the impacts human participants might have on the performance of an artificial market depending on the complexity of the task, e.g., there are some tasks which are very easy for humans but difficult for algorithms wherein we would expect near-perfect performance from human participants.

Our experiments measure impact of exogenous agents on market performance measured, as before, by accuracy, F1 score, and percentage of scored test points. Because exogenous agents are not trained, they are not subject to the genetic algorithm. Rather, exogenous agents are added directly to the agent pool during test. We test the impact of adding varying number of agents from each class. We specify this number based on percentage of the total agent pool. Specifically, we test hybrid market performance with the inclusion of $\{0.1\%,0.5\%,1\%\}$ GT and GTinv agents. We test hybrid market performance with the inclusion of random agents accounting for $\{1\%,5\%,10\%,50\%\}$ of the total agent pool. Random agents are included at a higher rate given the comparatively lesser impact they have on asset prices. All RQ2 experiments are run with an 80/20 train/test split. This process is diagrammed in Supplemental Materials.

All experiments with exogenous agents are based on the third of the three datasets studied in RQ1 for hyper-parameter assessment, namely, the replication outcomes data. This is the type of task where would expect the greatest gain from human-AI collaboration. Namely, this is an extremely challenging task for which (1) neither machine learning nor human judgement alone is likely to guarantee satisfactory performance, and for which (2) algorithmic and human assessments likely consider very different information/feature sets. 

\section{Results}

Following, we detail experimental findings in support of RQ1 and RQ2, respectively.

\subsection{Market robustness to hyper-parameters}

Market robustness to hyper-parameter settings is explored for the Iris and Heart Disease benchmark classification tasks, and for the prediction of replication studies outcomes. These experiments offer the opportunity to compare/contrast the impact of hyper-parameters across three contexts.

\subsubsection{Iris classification}

\begin{figure}[h]
    \centering
  \includegraphics[width=1\linewidth]{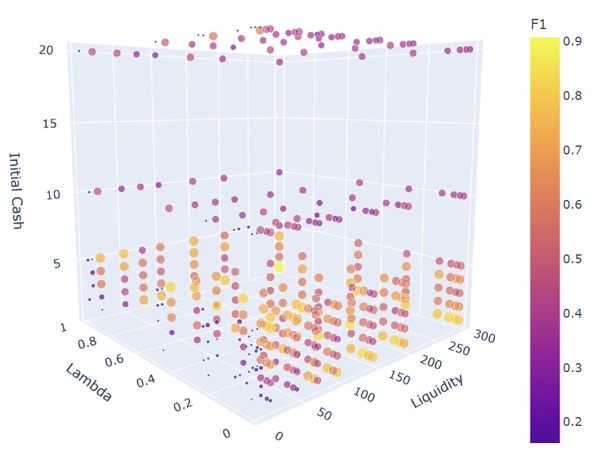}
  \caption{Average F1 score on the Iris classification task, plotted in hyper-parameter space.}
  \label{average_f1_4d_iris}
\end{figure}

Figure \ref{average_f1_4d_iris} highlights average F1 score over $441$ combinations of initial cash, $\lambda$, and liquidity factor. Generally, better F1 scores are obtained when initial cash ranges between $1$ and $4$ and when liquidity is greater than $100$. Choice of $\lambda$ does not appear to significantly impact F1 score. Best \textbf{F1} of \textbf{0.91} is achieved for \{liquidity factor $= 300$, $\lambda = 1.0$, initial cash $= 1$\}. In this setting, accuracy is $0.94$ and $100\%$ of the data is scored. 

\begin{figure}[h]
  \centering
  \includegraphics[width=0.95\linewidth]{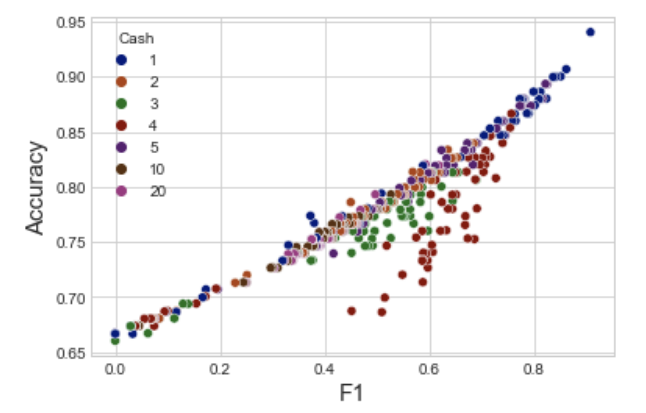}
  \caption{Average F1 score vs. accuracy on the Iris classification task, for varying initial cash.}
  \label{avg_acc_vs_f1_init_cash_iris}
\end{figure}

Tables and \ref{table_1_F1_Acc_initcash_iris}, \ref{table_2_F1_Acc_lambda_iris}, and \ref{f1_acc_liquidity_vary_iris} report market performance holding each one of the three hyper-parameters fixed and varying the other two. Performance metrics are averaged over $5$ folds. The data suggests that market performance increases as liquidity increases and decreases with initial cash. While, the effect of $\lambda$ reveals no clear pattern. 
Figure \ref{avg_acc_vs_f1_init_cash_iris} shows F1 vs. accuracy for different values of initial cash. F1 score increases with accuracy and best performance for both is achieved when initial cash is $1$. Similar plots of F1 vs. accuracy for liquidity and $\lambda$ are provided in Supplemental Materials.

\begin{table}[t]
 \caption{Average F1 score and accuracy on the Iris classification task, varying initial cash.}
\label{table_1_F1_Acc_initcash_iris}
\begin{tabular}{ccc|ccc}
\toprule
\textbf{Cash} &  Liquidity &  $\lambda$ &  Accuracy &  F1 & Scored \% \\
\midrule
\textbf{1}          &      300 &     1.0 &          0.94 &          0.91 & 100\\
\textbf{2}          &      300 &     1.0 &          0.81 &          0.58 & 100\\
\textbf{3}          &      300 &     1.0 &          0.81 &          0.64 & 100\\
\textbf{4}          &      300 &     1.0 &          0.87 &          0.76 & 100\\
\textbf{5}          &      300 &     1.0 &          0.76 &          0.42 & 100\\
\textbf{10}         &      300 &     1.0 &          0.75 &          0.35 & 100\\
\textbf{20}         &      300 &     1.0 &          0.75 &          0.37 & 100\\
\bottomrule
\end{tabular}
\end{table}

\begin{table}[t]
 \caption{Average F1 score and accuracy on the Iris classification task, varying $\lambda$.}
\label{table_2_F1_Acc_lambda_iris}
\begin{tabular}{ccc|ccc}
\toprule
\textbf{$\lambda$} &  Liquidity &   Cash &  Accuracy &  F1 & Scored \%\\
\midrule
\textbf{0.010}  &      300 &           1 &          0.87 &          0.79 & 100\\
\textbf{0.025}  &      300 &           1 &          0.91 &          0.86 & 100\\
\textbf{0.050}  &      300 &           1 &          0.88 &          0.82 & 100\\
\textbf{0.100}  &      300 &           1 &          0.87 &          0.79 & 100\\
\textbf{0.250}  &      300 &           1 &          0.87 &          0.80 & 100\\
\textbf{0.500}  &      300 &           1 &          0.87 &          0.78 & 100\\
\textbf{1.000}  &      300 &           1 &          0.94 &          0.91 & 100\\
\bottomrule
\end{tabular}
\end{table}

\begin{table}[t]
 \caption{Average F1 score and accuracy on the Iris classification task, varying liquidity.}
\label{f1_acc_liquidity_vary_iris}
\begin{tabular}{ccc|ccc}
\toprule
\textbf{Liquidity} &   $\lambda$ &   Cash &  Accuracy &  F1 & Scored \% \\
\midrule
\textbf{5}       &     1.0 &           1&          0.67 &          0.00 & 100 \\
\textbf{10}      &     1.0 &           1 &          0.67 &          0.00 & 100\\
\textbf{20}      &     1.0 &           1 &          0.67 &          0.03 & 100\\
\textbf{50}      &     1.0 &           1 &          0.75 &          0.33 & 100\\
\textbf{75}      &     1.0 &           1 &          0.85 &          0.72 & 100\\
\textbf{100}     &     1.0 &           1 &          0.86 &          0.74 & 100\\
\textbf{150}     &     1.0 &           1 &          0.86 &          0.76 & 100\\
\textbf{200}     &     1.0 &           1 &          0.88 &          0.81 & 100\\
\textbf{300}     &     1.0 &           1 &          0.94 &          0.91 & 100\\
\bottomrule
\end{tabular}
\end{table}

\subsubsection{Heart Disease classification}

\begin{figure}[h]
  \centering
  \includegraphics[width=1\linewidth]{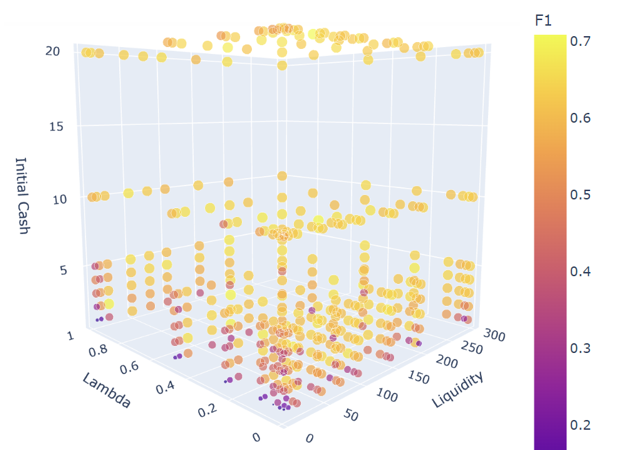}
  \caption{Average F1 score on the Heart Disease classification task, plotted in hyper-parameter space.}
    \label{average_f1_4d_heart}
\end{figure}

Figure \ref{average_f1_4d_heart} shows average F1 score for all $441$ hyper-parameter combinations. Performance is generally poorer than for the Iris classification task, and there is also not as clear a region of best performance in hyper-parameter space. Highest \textbf{F1} of \textbf{0.71} is achieved for (liquidity factor $= 50$, $\lambda = 0.05$, initial cash $= 20$). In this setting, accuracy is $0.66$ and $99.67\%$ of the data is scored (exactly one test point is left unscored by the market).

\begin{figure}[h]
  \centering
  \includegraphics[width=0.95\linewidth]{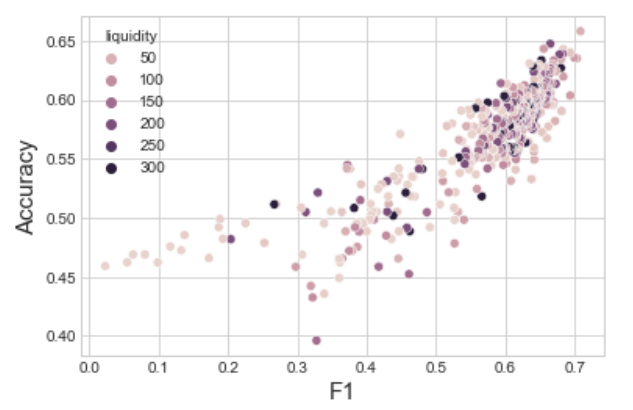}
  \caption{Average F1 score vs. accuracy on the Heart Disease classification task, for varying Liquidity}
  \label{avg_acc_vs_f1_liq_heart}
\end{figure}

As above, we report performance for varying liquidity in Table \ref{table_3_F1_Acc_liquidity_heart}, holding $\lambda$ and cash fixed. Similar tables are provided for varying $\lambda$ and cash in Supplemental Materials, as are plots of F1 vs. accuracy for $\lambda$, cash and liquidity. Similar to our finding on the Iris dataset, liquidity appears to be the primary driver of performance gains and losses on the Heart Disease classification task. Figure \ref{avg_acc_vs_f1_liq_heart} shows the average F1 vs. accuracy for all the combinations from where we took subsets to show more in-depth impact of liquidity. 

Of note, despite modest F1 and accuracy scores for this task, the percentage of scored test points is very high. This stands in contrast to results on the replication prediction task which follows. In other words,  agents in this case are sufficiently confident (have learned from sufficiently similar points in the training dataset) to invest. However, they are incorrect.  While, in the replication prediction task which follows, agents are not sufficiently confident and do not invest -- i.e., they ``know what they don't know''.

\begin{table}[t]
  \caption{Average F1 score and accuracy on the Heart Disease classification task, varying liquidity.}
  \label{table_3_F1_Acc_liquidity_heart}
  \begin{tabular}{ccc|ccc}\toprule
    \textbf{Liquidity} & $\lambda$ & Cash & Accuracy & F1 & Scored \% \\ 
    \midrule
    \textbf{5} & 0.05 & 20 & 0.59 & 0.54 & 100\\
    \textbf{10} & 0.05 & 20 & 0.54 & 0.56 & 99.67\\
    \textbf{20} & 0.05 & 20 & 0.60 & 0.66 & 99.01\\
    \textbf{50} & 0.05 & 20 & 0.66 & 0.71 & 99.67\\
    \textbf{75} & 0.05 & 20 & 0.61 & 0.64 & 99.67\\
    \textbf{100} & 0.05 & 20 & 0.58 & 0.62 & 99.34\\ 
    \textbf{150} & 0.05 & 20 & 0.58 & 0.59 & 99.01\\
    \textbf{200} & 0.05 & 20 & 0.57 & 0.62 & 99.34\\
    \textbf{300} & 0.05 & 20 & 0.59 & 0.61 & 99.34\\ 
    \bottomrule
  \end{tabular}
\end{table}


\subsubsection{Replication outcomes prediction}

\begin{figure}[h]
  \centering
  \includegraphics[width=1\linewidth]{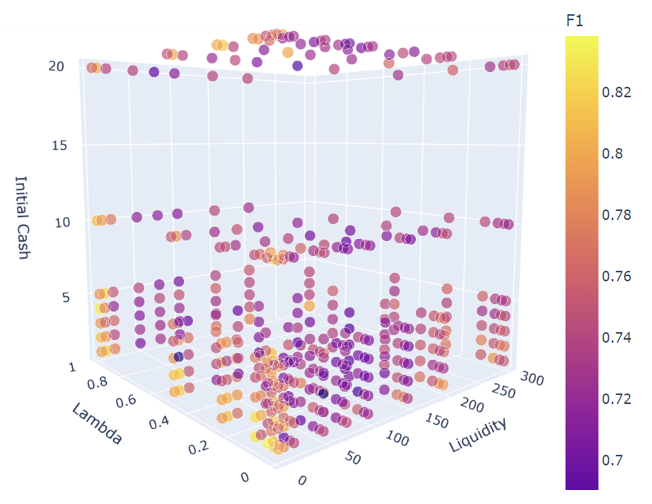}
  \caption{Average F1 score on the replication prediction task, plotted in hyper-parameter space.}
    \label{average_f1_4d_replication}
\end{figure}

\begin{figure}[h]
  \centering
  \includegraphics[width=0.95\linewidth]{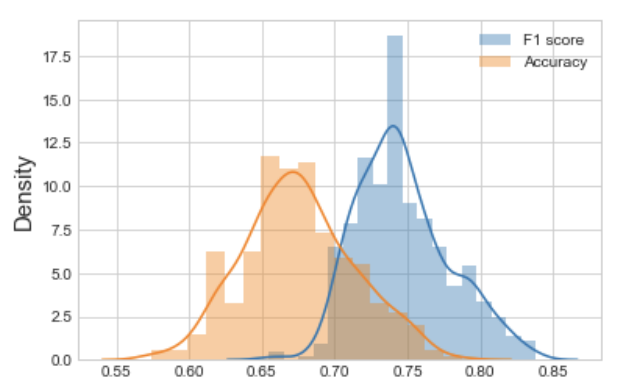}
  \caption{Density plot for F1 and accuracy scores on the replication prediction task.}
    \label{density_replication}
\end{figure}

\begin{figure}[h]
  \centering
  \includegraphics[width=0.95\linewidth]{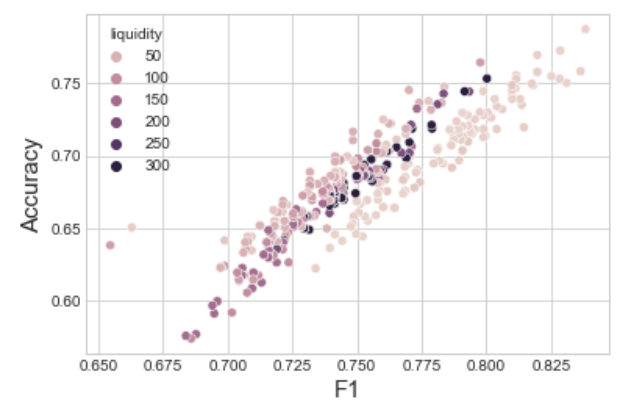}
  \caption{Average F1 score vs. accuracy with different values of Liquidity for Replication Data}
    \label{avg_acc_vs_f1_liquidity_replication}
\end{figure}

Finally, we explore the impact of hyper-parameter selection in the context of replication outcomes prediction. 
Figure \ref{average_f1_4d_replication} gives average F1 scores over all hyper-parameter combinations.  Best \textbf{F1} of \textbf{0.84} is achieved for \{liquidity factor $= 5$, $\lambda = 0.05$, initial cash $= 1$\}. In this setting, accuracy is $0.79$ and $36\%$ of the test data is scored.
Figure \ref{density_replication} provides another view of this data via the density plot of F1 and accuracy scores, across all hyper-parameter sets. Tables detailing F1, accuracy and percentage of scored test points, varying individual hyper-parameters, are provided in Supplemental Materials.

Figure \ref{avg_acc_vs_f1_liquidity_replication} shows the average F1 vs. accuracy scores on the replication prediction task, for varying liquidity. As was the case with both benchmark classification tasks, liquidity appears to drive performance here too. In this case, market performance improves as liquidity factor decreases. While, there are no clear best and worst-performing values for initial cash and $\lambda$. Supporting plots are shared in Supplemental Materials.

\begin{table}[t]
  \begin{tabular}{cc|cc|cc}\toprule
    Liquidity & \#Agents & Cash & \#Agents  & $\lambda$ &  \#Agents \\\midrule
5  &  24.49 & 1 & 35.39 & 0.01 & 35.08\\
10 &  26.15 & 2 & 35.87 & 0.025 & 34.93\\
20 &  29.13 & 3 & 35.22 & 0.05 & 35.06\\
50 &  35.43 & 4 & 34.96 & 0.1 & 35.13\\
75 &  37.48 & 5 & 34.79 & 0.25 & 35.18\\
100 & 38.88 & 10 & 34.90 & 0.5 & 35.06\\
150 & 40.57 & 20 & 34.73 & 1 & 35.41\\
200 & 41.51 & - & - & - & -\\
300 & 42.45 & - & - & - & -\\
\bottomrule
\end{tabular}
  \caption{Average number of agents participating in each market, of 1080 total agents, for varying liquidity, cash and $\lambda$, over all combinations of the other two parameters.}
  \label{table_agent_participation}
\end{table}

As noted, the artificial prediction market algorithm struggles with agent participation on the replication prediction task. The hyper-parameter set associated with highest F1 score leaves $63\%$ of the data unscored. In fact, all except two hyper-parameter combinations leave more than $40\%$ of the test data unscored (see Supplemental Materials). Liquidity factor here too appears to play a critical role. Table \ref{table_agent_participation} provides the average number of participating agents per market, for fixed values of each hyper-parameter. Liquidity has the greatest impact on participation. Liquidity controls the magnitude of shifts in asset price with each buy/sell. Agents' participation depends on movements in asset price, and as such, this behavior is in line with expectations.


\subsection{Market behavior with exogenous agents}

Our experiments in support of RQ2 introduce simulated, exogenous agents representing \emph{ground truth} (GT), \emph{ground truth inverse} (GTinv), and \emph{random} behavioral primitives into the market.  These additional agents operate outside of the training process and, as such, represent complementary actions that may underlie simple human participant inputs. Exogenous agents are introduced into the general agent pool and are subject to the same arrival rate, $\lambda$, as trained agents.

\begin{table}[t]
  \caption{Five best- and worst-performing hyper-parameter settings for replication prediction.}
  \label{table_10_best_worst}
  \begin{tabular}{ccc|ccc}\toprule
    Liquidity & $\lambda$ & Cash & F1 & Accuracy & Scored \% \\ \midrule
    5 & 0.05 & 1 & 0.84 & 0.79 & 36 \\
    10 & 0.05 & 10 & 0.84 & 0.76 & 35 \\
    5 & 1 & 4 & 0.83 & 0.75 & 37 \\
    5 & 0.1 & 1 & 0.83 & 0.77 & 36 \\
    5 & 0.1 & 2 & 0.83 & 0.75 & 37 \\ \midrule
    150 & 0.25 & 4 & 0.69 & 0.58 & 35 \\ 
    100 & 0.05 & 2 & 0.68 & 0.0.57 & 35 \\ 
    150 & 0.5 & 20 & 0.68 & 0.58 & 35 \\ 
    10 & 0.5 & 3 & 0.65 & 0.66 & 55 \\ 
    75 & 0.1 & 2 & 0.64 & 0.65 & 52 \\ \bottomrule
  \end{tabular}
\end{table}

Our simulations with exogenous agents are run over the replication prediction task as baseline. In particular, we consider the five best- and five worst-performing hyper-parameter settings, sorted by F1 score (Table \ref{table_10_best_worst}). We use these $10$ markets as baselines to study the impacts on performance of including GT, GTinv and random agents into the market.
Changes to F1 scores after the introduction of each of the three exogenous agent populations, in varying amounts, into the replication prediction markets are detailed in Table \ref{table_f1_exogenous}. Gains in accuracy follow similarly, see Supplemental Materials.

\begin{table*}[t]
  \caption{Average F1 scores on 10 replication prediction markets, for different types and size of exogenous agent populations}
  \label{table_f1_exogenous}
  \begin{tabular}{c|ccc|ccc|cccc}\toprule
    None & GT 0.1\% & GT 0.5\% & GT 1\% &  GTinv 0.1\% & GTinv 0.5\% & GTinv 1\% &  Random 1\% & Random 5\% & Random 10\% & Random 50\%\\ \midrule
    0.84 & 0.93 & 1 & 1 & 0.34 & 0.23 & 0.09 & 0.79 & 0.81 & 0.80 & 0.82 \\
    0.84 & 0.91 & 0.96 & 0.97 & 0.34 & 0.31 & 0.28 & 0.74 & 0.76 & 0.75 & 0.75 \\
    0.83 & 0.94 & 1 & 1 & 0.32 & 0.25 & 0.06 & 0.79 & 0.81 & 0.83 & 0.78 \\
    0.83 & 0.90 & 0.95 & 0.99 & 0.33 & 0.28 & 0.24 & 0.76 & 0.74 & 0.82 & 0.77 \\
    0.83 & 0.88 & 0.91 & 0.94 & 0.33 & 0.32 & 0.29 & 0.75 & 0.77 & 0.79 & 0.73 \\ \midrule
    0.69 & 0.89 & 0.94 & 0.96 & 0.34 & 0.31 & 0.28 & 0.76 & 0.77 & 0.81 & 0.77 \\ 
    0.69 & 0.91 & 0.97 & 0.96 & 0.34 & 0.29 & 0.29 & 0.78 & 0.77 & 0.77 & 0.78 \\ 
    0.68 & 0.90 & 0.95 & 0.96 & 0.33 & 0.29 & 0.29 & 0.78 & 0.78 & 0.79 & 0.76 \\ 
    0.66 & 0.89 & 0.90 & 0.94 & 0.33 & 0.33 & 0.29 & 0.76 & 0.76 & 0.79 & 0.74 \\ 
    0.65 & 0.90 & 0.91 & 0.94 & 0.34 & 0.32 & 0.30 & 0.77 & 0.75 & 0.81 & 0.74 \\ \bottomrule
  \end{tabular}
\end{table*}

\subsubsection{Introduction of GT agents into the agent pool}


\begin{figure}[h]
  \centering
  \includegraphics[width=0.95\linewidth]{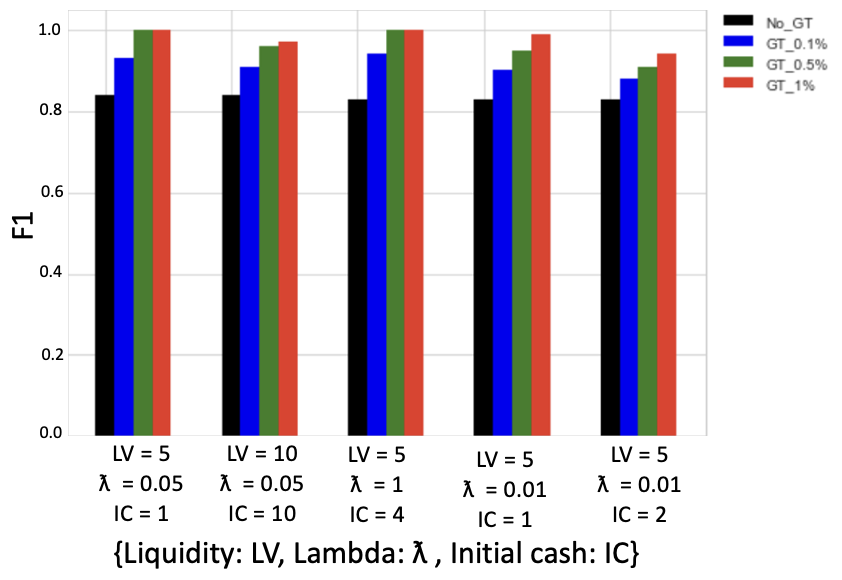}
  \caption{F1 with different percentage of added GT agents for replication data}
  \label{f1_gt_agents}
\end{figure}

Notably, the inclusion of even a very small population of GT agents improves market performance substantially. Figure \ref{f1_gt_agents} shows the incremental improvements in F1 score derived with as little as $0.1\%$ GT agents, i.e., $1$ GT agent for each $1000$ trained agents in the pool, for each of the five best hyper-parameter settings from the baseline replication outcomes experiments. Inclusion of $0.5\%$ GT agents brings F1 up to $1.0$ in two cases of the five, and over $0.9$ in all five.

\begin{figure}[h]
  \centering
  \includegraphics[width=0.95\linewidth]{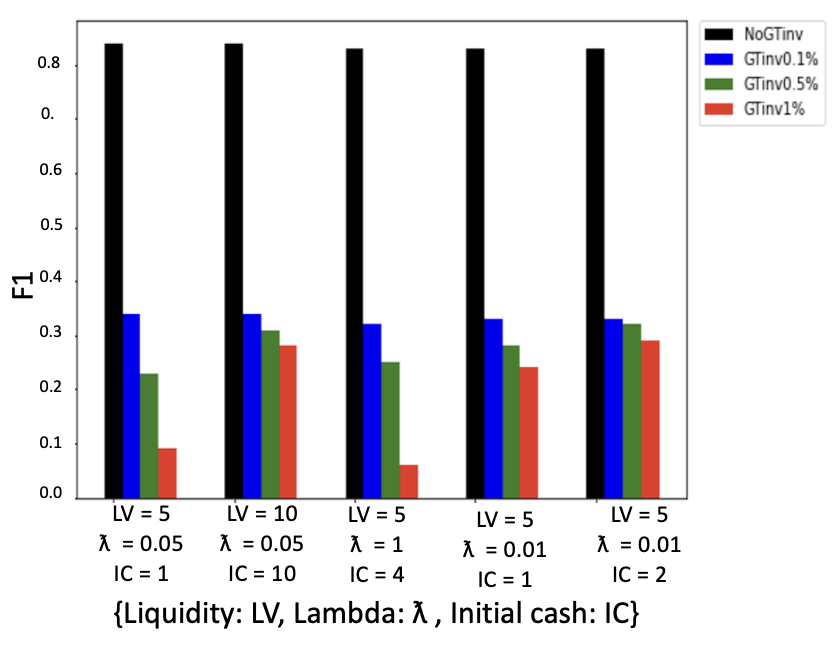}
  \caption{F1 with different percentage of added GT inverse agents for replication data}
  \label{f1_gt_inv_agents}
\end{figure}

\begin{table}[t]
  \caption{Agent participation with different percentages of GT and GTinv agents (1080 total agents).}
  \label{table_agent_part_gt_agents}
  \begin{tabular}{c|ccc|ccc}\toprule
    None & GT & GT & GT & GTinv & GTinv & GTinv\\ 
     & 0.1\% & 0.5\% & 1\% & 0.1\% & 0.5\% & 1\% \\
    \midrule
28.35 &    31.33 &    32.21 &  36.91 &       29.54 &       37.06 &     46.79\\
22.62 &    25.97 &    29.42 &  36.82 &       25.07 &       31.11 &     37.82\\
23.66 &    22.93 &    29.98 &  35.39 &       27.08 &       32.16 &     36.54\\
28.56 &    30.63 &    36.16 &  37.90 &       28.17 &       35.64 &     48.34\\
24.70 &    27.51 &    31.99 &  34.64 &       28.04 &       33.67 &     44.20\\\midrule
41.12 &    44.56 &    47.82 &  52.02 &       42.80 &       44.23 &     53.81\\
40.01 &    39.89 &    43.83 &  48.46 &       41.27 &       42.57 &     51.36\\
41.90 &    42.06 &    46.83 &  51.55 &       42.88 &       45.02 &     51.24\\
26.93 &    28.02 &    29.56 &  35.72 &       28.17 &       34.77 &     41.71\\
44.47 &    37.49 &    40.50 &  46.66 &       38.66 &       42.20 &     49.84\\
\bottomrule
\end{tabular}
\end{table}


\subsubsection{Introduction of GTinv agents into the agent pool}
The introduction of GTinv agents into the agent pool has even greater impact on F1, see Figure \ref{f1_gt_inv_agents}. The addition of just 0.1\% GTinv agents drops average F1 below 0.35 for all $10$ markets. While, inclusion of 1\% GTinv agents brings the best-performing baseline market from F1 of 0.84 to F1 of 0.09. 

The significant impact of very small numbers of GT and GTinv agents has important implications for the promise of future hybrid prediction markets with human participants. For a task which a human participant would very likely perform accurately but for which scaling is a concern, for example, this suggests that a trained, artificial prediction market might perform very well with minimal human input.

\subsubsection{Introduction of random agents into the agent pool}
Finally, and as an additional baseline, we experiment with the inclusion of agents who randomly buy and sell assets corresponding to future outcomes into the artifical prediction market framework. Because the impact of these agents is relatively lesser in magnitude than that of GT and GTinv agents, we experiment with adding more of them into the agent pool. In all cases, the inclusion of exogenous random agents into the agent pool degrades performance. However, in many cases, the change in performance is modest. These results are detailed in Supplemental Materials.

\subsubsection{Impact of exogenous agents on (trained) agent participation}

The inclusion of exogeneous agents into the agent pool has impact beyond their own asset purchases.  The investments of exogeneous agents in the market drives asset prices above and/or beyond where they were in the baseline case, and in doing so, has the impact of increasing participation amongst the trained agents in the pool. Table \ref{table_agent_part_gt_agents} gives the average number of participating trained agents, for each of the $10$ markets under study in RQ2, for each experimental condition. These trends are visualized in additional plots in Supplemental Materials. We note that increased participation is similarly observed for inclusion of GT and GTinv agents. Given the losses in F1 and accuracy when GTinv agents are present, it is clear that increased participation of trained agents is not necessarily a goal.  However, these findings highlight the possible impacts, both good and bad, of participation exogenous to the trained artificial market.

\section{Conclusions}

The comprehensive study of a simple artificial prediction market we undertake here highlights a promising new machine learning algorithm, which achieves respectable performance on benchmark machine learning tasks but which, we argue, affords unique opportunities for human-AI collaboration.

The performance of this very simple, initial market model is encouraging. There is likely great room for improvement: other agent training schemes may be more efficient that the genetic algorithm; more sophisticated agent logic can likely be devised; agents need not be homogeneous - rather, specialized agents populations may be trained with different and complementary expertise. These improvements, building on an already-functional baseline algorithm, may offer new avenues of creative artificial intelligence.

Beyond the potential of an artificial prediction market as an AI, future work should take the next step and introduce human participants into a hybrid prediction market model. This process will require research into best mechanisms and practices for human-agent collaboration in the context of markets. E.g., Should agents and human participants be given the same amount of cash?  What is the appropriate duration of such a market? At what rate should agents be permitted to transact? Which tasks are best suited to hybrid intelligence?  Ultimately, one goal might be to train a class of agents in the presence of human participants but be able to deploy those agents offline for scalability.

\bibliographystyle{ACM-Reference-Format} 
\bibliography{AAMASbib, RICObib, AAAIbib}


\begin{thebibliography}{79}


\ifx \showCODEN    \undefined \def \showCODEN     #1{\unskip}     \fi
\ifx \showDOI      \undefined \def \showDOI       #1{#1}\fi
\ifx \showISBNx    \undefined \def \showISBNx     #1{\unskip}     \fi
\ifx \showISBNxiii \undefined \def \showISBNxiii  #1{\unskip}     \fi
\ifx \showISSN     \undefined \def \showISSN      #1{\unskip}     \fi
\ifx \showLCCN     \undefined \def \showLCCN      #1{\unskip}     \fi
\ifx \shownote     \undefined \def \shownote      #1{#1}          \fi
\ifx \showarticletitle \undefined \def \showarticletitle #1{#1}   \fi
\ifx \showURL      \undefined \def \showURL       {\relax}        \fi
\providecommand\bibfield[2]{#2}
\providecommand\bibinfo[2]{#2}
\providecommand\natexlab[1]{#1}
\providecommand\showeprint[2][]{arXiv:#2}

\bibitem[\protect\citeauthoryear{Abernethy, Chen, and
  Wortman~Vaughan}{Abernethy et~al\mbox{.}}{2011}]%
        {abernethy2011optimization}
\bibfield{author}{\bibinfo{person}{Jacob Abernethy}, \bibinfo{person}{Yiling
  Chen}, {and} \bibinfo{person}{Jennifer Wortman~Vaughan}.}
  \bibinfo{year}{2011}\natexlab{}.
\newblock \showarticletitle{An optimization-based framework for automated
  market-making}. In \bibinfo{booktitle}{\emph{Proceedings of the 12th ACM
  conference on Electronic commerce}}. \bibinfo{pages}{297--306}.
\newblock


\bibitem[\protect\citeauthoryear{Altmejd, Dreber, Forsell, Huber, Imai,
  Johannesson, Kirchler, Nave, and Camerer}{Altmejd et~al\mbox{.}}{2019}]%
        {altmejd2019predicting}
\bibfield{author}{\bibinfo{person}{Adam Altmejd}, \bibinfo{person}{Anna
  Dreber}, \bibinfo{person}{Eskil Forsell}, \bibinfo{person}{Juergen Huber},
  \bibinfo{person}{Taisuke Imai}, \bibinfo{person}{Magnus Johannesson},
  \bibinfo{person}{Michael Kirchler}, \bibinfo{person}{Gideon Nave}, {and}
  \bibinfo{person}{Colin Camerer}.} \bibinfo{year}{2019}\natexlab{}.
\newblock \showarticletitle{Predicting the replicability of social science lab
  experiments}.
\newblock \bibinfo{journal}{\emph{PloS one}} \bibinfo{volume}{14},
  \bibinfo{number}{12} (\bibinfo{year}{2019}), \bibinfo{pages}{e0225826}.
\newblock


\bibitem[\protect\citeauthoryear{Amershi, Cakmak, Knox, and Kulesza}{Amershi
  et~al\mbox{.}}{2014}]%
        {amershi2014power}
\bibfield{author}{\bibinfo{person}{Saleema Amershi}, \bibinfo{person}{Maya
  Cakmak}, \bibinfo{person}{William~Bradley Knox}, {and} \bibinfo{person}{Todd
  Kulesza}.} \bibinfo{year}{2014}\natexlab{}.
\newblock \showarticletitle{Power to the people: The role of humans in
  interactive machine learning}.
\newblock \bibinfo{journal}{\emph{Ai Magazine}} \bibinfo{volume}{35},
  \bibinfo{number}{4} (\bibinfo{year}{2014}), \bibinfo{pages}{105--120}.
\newblock


\bibitem[\protect\citeauthoryear{Amershi, Weld, Vorvoreanu, Fourney, Nushi,
  Collisson, Suh, Iqbal, Bennett, Inkpen, et~al\mbox{.}}{Amershi
  et~al\mbox{.}}{2019}]%
        {amershi2019guidelines}
\bibfield{author}{\bibinfo{person}{Saleema Amershi}, \bibinfo{person}{Dan
  Weld}, \bibinfo{person}{Mihaela Vorvoreanu}, \bibinfo{person}{Adam Fourney},
  \bibinfo{person}{Besmira Nushi}, \bibinfo{person}{Penny Collisson},
  \bibinfo{person}{Jina Suh}, \bibinfo{person}{Shamsi Iqbal},
  \bibinfo{person}{Paul~N Bennett}, \bibinfo{person}{Kori Inkpen},
  {et~al\mbox{.}}} \bibinfo{year}{2019}\natexlab{}.
\newblock \showarticletitle{Guidelines for human-AI interaction}. In
  \bibinfo{booktitle}{\emph{Proceedings of the 2019 chi conference on human
  factors in computing systems}}. \bibinfo{pages}{1--13}.
\newblock


\bibitem[\protect\citeauthoryear{Arrow, Forsythe, Gorham, Hahn, Hanson,
  Ledyard, Levmore, Litan, Milgrom, Nelson, Neumann, Ottaviani, Schelling,
  Shiller, Smith, Snowberg, Sunstein, Tetlock, Tetlock, Varian, Wolfers, and
  Zitzewitz}{Arrow et~al\mbox{.}}{2008}]%
        {Arrow:2008}
\bibfield{author}{\bibinfo{person}{Kenneth~J. Arrow}, \bibinfo{person}{Robert
  Forsythe}, \bibinfo{person}{Michael Gorham}, \bibinfo{person}{Robert Hahn},
  \bibinfo{person}{Robin Hanson}, \bibinfo{person}{John~O. Ledyard},
  \bibinfo{person}{Saul Levmore}, \bibinfo{person}{Robert Litan},
  \bibinfo{person}{Paul Milgrom}, \bibinfo{person}{Forrest~D. Nelson},
  \bibinfo{person}{George~R. Neumann}, \bibinfo{person}{Marco Ottaviani},
  \bibinfo{person}{Thomas~C. Schelling}, \bibinfo{person}{Robert~J. Shiller},
  \bibinfo{person}{Vernon~L. Smith}, \bibinfo{person}{Erik Snowberg},
  \bibinfo{person}{Cass~R. Sunstein}, \bibinfo{person}{Paul~C. Tetlock},
  \bibinfo{person}{Philip~E. Tetlock}, \bibinfo{person}{Hal~R. Varian},
  \bibinfo{person}{Justin Wolfers}, {and} \bibinfo{person}{Eric Zitzewitz}.}
  \bibinfo{year}{2008}\natexlab{}.
\newblock \showarticletitle{The Promise of Prediction Markets}.
\newblock \bibinfo{journal}{\emph{Science}} \bibinfo{volume}{320},
  \bibinfo{number}{5878} (\bibinfo{date}{May} \bibinfo{year}{2008}),
  \bibinfo{pages}{877--878}.
\newblock


\bibitem[\protect\citeauthoryear{Bansal, Nushi, Kamar, Lasecki, Weld, and
  Horvitz}{Bansal et~al\mbox{.}}{2019}]%
        {bansal2019beyond}
\bibfield{author}{\bibinfo{person}{Gagan Bansal}, \bibinfo{person}{Besmira
  Nushi}, \bibinfo{person}{Ece Kamar}, \bibinfo{person}{Walter~S Lasecki},
  \bibinfo{person}{Daniel~S Weld}, {and} \bibinfo{person}{Eric Horvitz}.}
  \bibinfo{year}{2019}\natexlab{}.
\newblock \showarticletitle{Beyond accuracy: The role of mental models in
  human-AI team performance}. In \bibinfo{booktitle}{\emph{Proceedings of the
  AAAI Conference on Human Computation and Crowdsourcing}},
  Vol.~\bibinfo{volume}{7}. \bibinfo{pages}{2--11}.
\newblock


\bibitem[\protect\citeauthoryear{Barbu and Lay}{Barbu and Lay}{2013}]%
        {barbu2013artificial}
\bibfield{author}{\bibinfo{person}{Adrian Barbu} {and} \bibinfo{person}{Nathan
  Lay}.} \bibinfo{year}{2013}\natexlab{}.
\newblock \showarticletitle{Artificial prediction markets for lymph node
  detection}. In \bibinfo{booktitle}{\emph{2013 E-Health and Bioengineering
  Conference (EHB)}}. IEEE, \bibinfo{pages}{1--7}.
\newblock


\bibitem[\protect\citeauthoryear{Barbu, Lay, and Mannor}{Barbu
  et~al\mbox{.}}{2012}]%
        {barbu2012introduction}
\bibfield{author}{\bibinfo{person}{Adrian Barbu}, \bibinfo{person}{Nathan Lay},
  {and} \bibinfo{person}{Shie Mannor}.} \bibinfo{year}{2012}\natexlab{}.
\newblock \showarticletitle{An Introduction to Artificial Prediction Markets
  for Classification.}
\newblock \bibinfo{journal}{\emph{Journal of Machine Learning Research}}
  \bibinfo{volume}{13}, \bibinfo{number}{7} (\bibinfo{year}{2012}).
\newblock


\bibitem[\protect\citeauthoryear{Berg, Nelson, and Rietz}{Berg
  et~al\mbox{.}}{2008}]%
        {berg2008prediction}
\bibfield{author}{\bibinfo{person}{Joyce~E Berg}, \bibinfo{person}{Forrest~D
  Nelson}, {and} \bibinfo{person}{Thomas~A Rietz}.}
  \bibinfo{year}{2008}\natexlab{}.
\newblock \showarticletitle{Prediction market accuracy in the long run}.
\newblock \bibinfo{journal}{\emph{International Journal of Forecasting}}
  \bibinfo{volume}{24}, \bibinfo{number}{2} (\bibinfo{year}{2008}),
  \bibinfo{pages}{285--300}.
\newblock


\bibitem[\protect\citeauthoryear{Brown and Sandholm}{Brown and
  Sandholm}{2019}]%
        {brown2019superhuman}
\bibfield{author}{\bibinfo{person}{Noam Brown} {and} \bibinfo{person}{Tuomas
  Sandholm}.} \bibinfo{year}{2019}\natexlab{}.
\newblock \showarticletitle{Superhuman AI for multiplayer poker}.
\newblock \bibinfo{journal}{\emph{Science}} \bibinfo{volume}{365},
  \bibinfo{number}{6456} (\bibinfo{year}{2019}), \bibinfo{pages}{885--890}.
\newblock


\bibitem[\protect\citeauthoryear{Camerer, Dreber, Forsell, Ho, Huber,
  Johannesson, Kirchler, Almenberg, Altmejd, Chan, et~al\mbox{.}}{Camerer
  et~al\mbox{.}}{2016}]%
        {camerer2016evaluating}
\bibfield{author}{\bibinfo{person}{Colin~F Camerer}, \bibinfo{person}{Anna
  Dreber}, \bibinfo{person}{Eskil Forsell}, \bibinfo{person}{Teck-Hua Ho},
  \bibinfo{person}{J{\"u}rgen Huber}, \bibinfo{person}{Magnus Johannesson},
  \bibinfo{person}{Michael Kirchler}, \bibinfo{person}{Johan Almenberg},
  \bibinfo{person}{Adam Altmejd}, \bibinfo{person}{Taizan Chan},
  {et~al\mbox{.}}} \bibinfo{year}{2016}\natexlab{}.
\newblock \showarticletitle{Evaluating replicability of laboratory experiments
  in economics}.
\newblock \bibinfo{journal}{\emph{Science}} \bibinfo{volume}{351},
  \bibinfo{number}{6280} (\bibinfo{year}{2016}), \bibinfo{pages}{1433--1436}.
\newblock


\bibitem[\protect\citeauthoryear{Camerer, Dreber, Holzmeister, Ho, Huber,
  Johannesson, Kirchler, Nave, Nosek, Pfeiffer, et~al\mbox{.}}{Camerer
  et~al\mbox{.}}{2018}]%
        {camerer2018evaluating}
\bibfield{author}{\bibinfo{person}{Colin~F Camerer}, \bibinfo{person}{Anna
  Dreber}, \bibinfo{person}{Felix Holzmeister}, \bibinfo{person}{Teck-Hua Ho},
  \bibinfo{person}{J{\"u}rgen Huber}, \bibinfo{person}{Magnus Johannesson},
  \bibinfo{person}{Michael Kirchler}, \bibinfo{person}{Gideon Nave},
  \bibinfo{person}{Brian~A Nosek}, \bibinfo{person}{Thomas Pfeiffer},
  {et~al\mbox{.}}} \bibinfo{year}{2018}\natexlab{}.
\newblock \showarticletitle{Evaluating the replicability of social science
  experiments in Nature and Science between 2010 and 2015}.
\newblock \bibinfo{journal}{\emph{Nature Human Behaviour}} \bibinfo{volume}{2},
  \bibinfo{number}{9} (\bibinfo{year}{2018}), \bibinfo{pages}{637--644}.
\newblock
\urldef\tempurl%
\url{https://doi.org/10.1038/s41562-018-0399-z}
\showDOI{\tempurl}


\bibitem[\protect\citeauthoryear{Canonico, Flathmann, and McNeese}{Canonico
  et~al\mbox{.}}{2019}]%
        {canonico2019wisdom}
\bibfield{author}{\bibinfo{person}{Lorenzo~Barberis Canonico},
  \bibinfo{person}{Christopher Flathmann}, {and} \bibinfo{person}{Nathan
  McNeese}.} \bibinfo{year}{2019}\natexlab{}.
\newblock \showarticletitle{The wisdom of the market: Using human factors to
  design prediction markets for collective intelligence}. In
  \bibinfo{booktitle}{\emph{Proceedings of the Human Factors and Ergonomics
  Society Annual Meeting}}, Vol.~\bibinfo{volume}{63}. SAGE Publications Sage
  CA: Los Angeles, CA, \bibinfo{pages}{1471--1475}.
\newblock


\bibitem[\protect\citeauthoryear{Chen, Fortnow, Lambert, Pennock, and
  Wortman}{Chen et~al\mbox{.}}{2008}]%
        {chen2008complexity}
\bibfield{author}{\bibinfo{person}{Yiling Chen}, \bibinfo{person}{Lance
  Fortnow}, \bibinfo{person}{Nicolas Lambert}, \bibinfo{person}{David~M
  Pennock}, {and} \bibinfo{person}{Jennifer Wortman}.}
  \bibinfo{year}{2008}\natexlab{}.
\newblock \showarticletitle{Complexity of combinatorial market makers}. In
  \bibinfo{booktitle}{\emph{Proceedings of the 9th ACM Conference on Electronic
  Commerce}}. \bibinfo{pages}{190--199}.
\newblock


\bibitem[\protect\citeauthoryear{Chen and Pennock}{Chen and Pennock}{[n.d.]}]%
        {ChenPennock:2010}
\bibfield{author}{\bibinfo{person}{Yiling Chen} {and} \bibinfo{person}{David~M.
  Pennock}.} \bibinfo{year}{[n.d.]}\natexlab{}.
\newblock \showarticletitle{Designing markets for prediction}.
\newblock \bibinfo{journal}{\emph{{AI} Magazine}} \bibinfo{volume}{31},
  \bibinfo{number}{4} (\bibinfo{year}{[n.\,d.]}), \bibinfo{pages}{42--–52}.
\newblock


\bibitem[\protect\citeauthoryear{Chen and Vaughan}{Chen and Vaughan}{2010}]%
        {chen2010new}
\bibfield{author}{\bibinfo{person}{Yiling Chen} {and}
  \bibinfo{person}{Jennifer~Wortman Vaughan}.} \bibinfo{year}{2010}\natexlab{}.
\newblock \showarticletitle{A new understanding of prediction markets via
  no-regret learning}. In \bibinfo{booktitle}{\emph{Proceedings of the 11th ACM
  conference on Electronic commerce}}. \bibinfo{pages}{189--198}.
\newblock


\bibitem[\protect\citeauthoryear{Chicho, Abdulazeez, Zeebaree, and
  Zebari}{Chicho et~al\mbox{.}}{2021}]%
        {chicho2021machine}
\bibfield{author}{\bibinfo{person}{Bahzad~Taha Chicho},
  \bibinfo{person}{Adnan~Mohsin Abdulazeez}, \bibinfo{person}{Diyar~Qader
  Zeebaree}, {and} \bibinfo{person}{Dilovan~Assad Zebari}.}
  \bibinfo{year}{2021}\natexlab{}.
\newblock \showarticletitle{Machine learning classifiers based classification
  for IRIS recognition}.
\newblock \bibinfo{journal}{\emph{Qubahan Academic Journal}}
  \bibinfo{volume}{1}, \bibinfo{number}{2} (\bibinfo{year}{2021}),
  \bibinfo{pages}{106--118}.
\newblock


\bibitem[\protect\citeauthoryear{Cova, Strickland, Abatista, Allard, Andow,
  Attie, Beebe, Berni{\=u}nas, Boudesseul, Colombo, et~al\mbox{.}}{Cova
  et~al\mbox{.}}{2021}]%
        {cova2021estimating}
\bibfield{author}{\bibinfo{person}{Florian Cova}, \bibinfo{person}{Brent
  Strickland}, \bibinfo{person}{Angela Abatista}, \bibinfo{person}{Aur{\'e}lien
  Allard}, \bibinfo{person}{James Andow}, \bibinfo{person}{Mario Attie},
  \bibinfo{person}{James Beebe}, \bibinfo{person}{Renatas Berni{\=u}nas},
  \bibinfo{person}{Jordane Boudesseul}, \bibinfo{person}{Matteo Colombo},
  {et~al\mbox{.}}} \bibinfo{year}{2021}\natexlab{}.
\newblock \showarticletitle{Estimating the reproducibility of experimental
  philosophy}.
\newblock \bibinfo{journal}{\emph{Review of Philosophy and Psychology}}
  \bibinfo{volume}{12}, \bibinfo{number}{1} (\bibinfo{year}{2021}),
  \bibinfo{pages}{9--44}.
\newblock


\bibitem[\protect\citeauthoryear{Cowgill, Wolfers, and Zitzewitz}{Cowgill
  et~al\mbox{.}}{2009}]%
        {cowgill2009using}
\bibfield{author}{\bibinfo{person}{Bo Cowgill}, \bibinfo{person}{Justin
  Wolfers}, {and} \bibinfo{person}{Eric Zitzewitz}.}
  \bibinfo{year}{2009}\natexlab{}.
\newblock \showarticletitle{Using Prediction Markets to Track Information
  Flows: Evidence from Google.}. In \bibinfo{booktitle}{\emph{amma}}.
  \bibinfo{pages}{3}.
\newblock


\bibitem[\protect\citeauthoryear{Dellermann, Calma, Lipusch, Weber, Weigel, and
  Ebel}{Dellermann et~al\mbox{.}}{2021}]%
        {dellermann2021future}
\bibfield{author}{\bibinfo{person}{Dominik Dellermann}, \bibinfo{person}{Adrian
  Calma}, \bibinfo{person}{Nikolaus Lipusch}, \bibinfo{person}{Thorsten Weber},
  \bibinfo{person}{Sascha Weigel}, {and} \bibinfo{person}{Philipp Ebel}.}
  \bibinfo{year}{2021}\natexlab{}.
\newblock \showarticletitle{The future of human-AI collaboration: a taxonomy of
  design knowledge for hybrid intelligence systems}.
\newblock \bibinfo{journal}{\emph{arXiv preprint arXiv:2105.03354}}
  (\bibinfo{year}{2021}).
\newblock


\bibitem[\protect\citeauthoryear{Desai, Giraddi, Narayankar, Pudakalakatti, and
  Sulegaon}{Desai et~al\mbox{.}}{2019}]%
        {desai2019back}
\bibfield{author}{\bibinfo{person}{Shrinivas~D Desai},
  \bibinfo{person}{Shantala Giraddi}, \bibinfo{person}{Prashant Narayankar},
  \bibinfo{person}{Neha~R Pudakalakatti}, {and} \bibinfo{person}{Shreya
  Sulegaon}.} \bibinfo{year}{2019}\natexlab{}.
\newblock \showarticletitle{Back-propagation neural network versus logistic
  regression in heart disease classification}.
\newblock In \bibinfo{booktitle}{\emph{Advanced computing and communication
  technologies}}. \bibinfo{publisher}{Springer}, \bibinfo{pages}{133--144}.
\newblock


\bibitem[\protect\citeauthoryear{Dreber, Pfeiffer, Almenberg, Isaksson, Wilson,
  Chen, Nosek, and Johannesson}{Dreber et~al\mbox{.}}{2015}]%
        {dreber2015using}
\bibfield{author}{\bibinfo{person}{Anna Dreber}, \bibinfo{person}{Thomas
  Pfeiffer}, \bibinfo{person}{Johan Almenberg}, \bibinfo{person}{Siri
  Isaksson}, \bibinfo{person}{Brad Wilson}, \bibinfo{person}{Yiling Chen},
  \bibinfo{person}{Brian~A Nosek}, {and} \bibinfo{person}{Magnus Johannesson}.}
  \bibinfo{year}{2015}\natexlab{}.
\newblock \showarticletitle{Using prediction markets to estimate the
  reproducibility of scientific research}.
\newblock \bibinfo{journal}{\emph{Proceedings of the National Academy of
  Sciences}} \bibinfo{volume}{112}, \bibinfo{number}{50}
  (\bibinfo{year}{2015}), \bibinfo{pages}{15343--15347}.
\newblock


\bibitem[\protect\citeauthoryear{Errington, Iorns, Gunn, Tan, Lomax, and
  Nosek}{Errington et~al\mbox{.}}{2014}]%
        {errington2014open}
\bibfield{author}{\bibinfo{person}{Timothy~M Errington},
  \bibinfo{person}{Elizabeth Iorns}, \bibinfo{person}{William Gunn},
  \bibinfo{person}{Fraser~Elisabeth Tan}, \bibinfo{person}{Joelle Lomax}, {and}
  \bibinfo{person}{Brian~A Nosek}.} \bibinfo{year}{2014}\natexlab{}.
\newblock \showarticletitle{An open investigation of the reproducibility of
  cancer biology research}.
\newblock \bibinfo{journal}{\emph{Elife}}  \bibinfo{volume}{3}
  (\bibinfo{year}{2014}).
\newblock


\bibitem[\protect\citeauthoryear{Fisher}{Fisher}{1988}]%
        {misc_iris_53}
\bibfield{author}{\bibinfo{person}{R.A. Fisher}.}
  \bibinfo{year}{1988}\natexlab{}.
\newblock \bibinfo{title}{{Iris}}.
\newblock \bibinfo{howpublished}{UCI Machine Learning Repository}.
\newblock


\bibitem[\protect\citeauthoryear{Fogliato, De-Arteaga, and
  Chouldechova}{Fogliato et~al\mbox{.}}{2022}]%
        {fogliato2022case}
\bibfield{author}{\bibinfo{person}{Riccardo Fogliato}, \bibinfo{person}{Maria
  De-Arteaga}, {and} \bibinfo{person}{Alexandra Chouldechova}.}
  \bibinfo{year}{2022}\natexlab{}.
\newblock \showarticletitle{A Case for Humans-in-the-Loop: Decisions in the
  Presence of Misestimated Algorithmic Scores}.
\newblock \bibinfo{journal}{\emph{Available at SSRN 4050125}}
  (\bibinfo{year}{2022}).
\newblock


\bibitem[\protect\citeauthoryear{Forsell, Viganola, Pfeiffer, Almenberg,
  Wilson, Chen, Nosek, Johannesson, and Dreber}{Forsell et~al\mbox{.}}{2019}]%
        {forsell2019predicting}
\bibfield{author}{\bibinfo{person}{Eskil Forsell}, \bibinfo{person}{Domenico
  Viganola}, \bibinfo{person}{Thomas Pfeiffer}, \bibinfo{person}{Johan
  Almenberg}, \bibinfo{person}{Brad Wilson}, \bibinfo{person}{Yiling Chen},
  \bibinfo{person}{Brian~A Nosek}, \bibinfo{person}{Magnus Johannesson}, {and}
  \bibinfo{person}{Anna Dreber}.} \bibinfo{year}{2019}\natexlab{}.
\newblock \showarticletitle{Predicting replication outcomes in the Many Labs 2
  study}.
\newblock \bibinfo{journal}{\emph{Journal of Economic Psychology}}
  \bibinfo{volume}{75} (\bibinfo{year}{2019}), \bibinfo{pages}{102117}.
\newblock


\bibitem[\protect\citeauthoryear{Gillen, Plott, and Shum}{Gillen
  et~al\mbox{.}}{2012}]%
        {gillen2012information}
\bibfield{author}{\bibinfo{person}{Benjamin~J Gillen},
  \bibinfo{person}{Charles~R Plott}, {and} \bibinfo{person}{Matthew Shum}.}
  \bibinfo{year}{2012}\natexlab{}.
\newblock \showarticletitle{Information aggregation mechanisms in the field:
  Sales forecasting inside intel}.
\newblock In \bibinfo{booktitle}{\emph{Working paper}}.
\newblock


\bibitem[\protect\citeauthoryear{Gordon, Viganola, Bishop, Chen, Dreber,
  Goldfedder, Holzmeister, Johannesson, Liu, Twardy, et~al\mbox{.}}{Gordon
  et~al\mbox{.}}{2020}]%
        {gordon2020replication}
\bibfield{author}{\bibinfo{person}{Michael Gordon}, \bibinfo{person}{Domenico
  Viganola}, \bibinfo{person}{Michael Bishop}, \bibinfo{person}{Yiling Chen},
  \bibinfo{person}{Anna Dreber}, \bibinfo{person}{Brandon Goldfedder},
  \bibinfo{person}{Felix Holzmeister}, \bibinfo{person}{Magnus Johannesson},
  \bibinfo{person}{Yang Liu}, \bibinfo{person}{Charles Twardy},
  {et~al\mbox{.}}} \bibinfo{year}{2020}\natexlab{}.
\newblock \showarticletitle{Are replication rates the same across academic
  fields? Community forecasts from the DARPA SCORE programme}.
\newblock \bibinfo{journal}{\emph{Royal Society open science}}
  (\bibinfo{year}{2020}).
\newblock


\bibitem[\protect\citeauthoryear{Gordon, Viganola, Dreber, Johannesson, and
  Pfeiffer}{Gordon et~al\mbox{.}}{2021}]%
        {gordon2021predicting}
\bibfield{author}{\bibinfo{person}{Michael Gordon}, \bibinfo{person}{Domenico
  Viganola}, \bibinfo{person}{Anna Dreber}, \bibinfo{person}{Magnus
  Johannesson}, {and} \bibinfo{person}{Thomas Pfeiffer}.}
  \bibinfo{year}{2021}\natexlab{}.
\newblock \showarticletitle{Predicting replicability—Analysis of survey and
  prediction market data from large-scale forecasting projects}.
\newblock \bibinfo{journal}{\emph{Plos one}} \bibinfo{volume}{16},
  \bibinfo{number}{4} (\bibinfo{year}{2021}), \bibinfo{pages}{e0248780}.
\newblock


\bibitem[\protect\citeauthoryear{Green and Chen}{Green and Chen}{2019}]%
        {green2019principles}
\bibfield{author}{\bibinfo{person}{Ben Green} {and} \bibinfo{person}{Yiling
  Chen}.} \bibinfo{year}{2019}\natexlab{}.
\newblock \showarticletitle{The principles and limits of algorithm-in-the-loop
  decision making}.
\newblock \bibinfo{journal}{\emph{Proceedings of the ACM on Human-Computer
  Interaction}} \bibinfo{volume}{3}, \bibinfo{number}{CSCW}
  (\bibinfo{year}{2019}), \bibinfo{pages}{1--24}.
\newblock


\bibitem[\protect\citeauthoryear{Hanson, Oprea, and Porter}{Hanson
  et~al\mbox{.}}{2006}]%
        {hanson2006information}
\bibfield{author}{\bibinfo{person}{Robin Hanson}, \bibinfo{person}{Ryan Oprea},
  {and} \bibinfo{person}{David Porter}.} \bibinfo{year}{2006}\natexlab{}.
\newblock \showarticletitle{Information aggregation and manipulation in an
  experimental market}.
\newblock \bibinfo{journal}{\emph{Journal of Economic Behavior \&
  Organization}} \bibinfo{volume}{60}, \bibinfo{number}{4}
  (\bibinfo{year}{2006}), \bibinfo{pages}{449--459}.
\newblock


\bibitem[\protect\citeauthoryear{Harper}{Harper}{2019}]%
        {harper2019role}
\bibfield{author}{\bibinfo{person}{Richard~HR Harper}.}
  \bibinfo{year}{2019}\natexlab{}.
\newblock \showarticletitle{The Role of HCI in the Age of AI}.
\newblock \bibinfo{journal}{\emph{International Journal of Human--Computer
  Interaction}} \bibinfo{volume}{35}, \bibinfo{number}{15}
  (\bibinfo{year}{2019}), \bibinfo{pages}{1331--1344}.
\newblock


\bibitem[\protect\citeauthoryear{He, Zhang, Ren, and Sun}{He
  et~al\mbox{.}}{2015}]%
        {he2015delving}
\bibfield{author}{\bibinfo{person}{Kaiming He}, \bibinfo{person}{Xiangyu
  Zhang}, \bibinfo{person}{Shaoqing Ren}, {and} \bibinfo{person}{Jian Sun}.}
  \bibinfo{year}{2015}\natexlab{}.
\newblock \showarticletitle{Delving deep into rectifiers: Surpassing
  human-level performance on imagenet classification}. In
  \bibinfo{booktitle}{\emph{Proceedings of the IEEE international conference on
  computer vision}}. \bibinfo{pages}{1026--1034}.
\newblock


\bibitem[\protect\citeauthoryear{Hu and Storkey}{Hu and Storkey}{2014}]%
        {hu2014multi}
\bibfield{author}{\bibinfo{person}{Jinli Hu} {and} \bibinfo{person}{Amos
  Storkey}.} \bibinfo{year}{2014}\natexlab{}.
\newblock \showarticletitle{Multi-period trading prediction markets with
  connections to machine learning}. In \bibinfo{booktitle}{\emph{International
  Conference on Machine Learning}}. \bibinfo{pages}{1773--1781}.
\newblock


\bibitem[\protect\citeauthoryear{Jahedpari, Padget, De~Vos, and
  Hirsch}{Jahedpari et~al\mbox{.}}{2014}]%
        {jahedpari2014artificial}
\bibfield{author}{\bibinfo{person}{Fatemeh Jahedpari}, \bibinfo{person}{Julian
  Padget}, \bibinfo{person}{Marina De~Vos}, {and} \bibinfo{person}{Benjamin
  Hirsch}.} \bibinfo{year}{2014}\natexlab{}.
\newblock \showarticletitle{Artificial prediction markets as a tool for
  syndromic surveillance}.
\newblock \bibinfo{journal}{\emph{Crowd Intelligence: Foundations, Methods and
  Practices}} (\bibinfo{year}{2014}).
\newblock


\bibitem[\protect\citeauthoryear{Janosi, Steinbrunn, Pfisterer, and
  Detrano}{Janosi et~al\mbox{.}}{1988}]%
        {misc_heart_disease_45}
\bibfield{author}{\bibinfo{person}{Andras Janosi}, \bibinfo{person}{William
  Steinbrunn}, \bibinfo{person}{Matthias Pfisterer}, {and}
  \bibinfo{person}{Robert Detrano}.} \bibinfo{year}{1988}\natexlab{}.
\newblock \showarticletitle{Heart disease data set}.
\newblock \bibinfo{journal}{\emph{The UCI KDD Archive}} (\bibinfo{year}{1988}).
\newblock


\bibitem[\protect\citeauthoryear{Jarrahi}{Jarrahi}{2018}]%
        {jarrahi2018artificial}
\bibfield{author}{\bibinfo{person}{Mohammad~Hossein Jarrahi}.}
  \bibinfo{year}{2018}\natexlab{}.
\newblock \showarticletitle{Artificial intelligence and the future of work:
  Human-AI symbiosis in organizational decision making}.
\newblock \bibinfo{journal}{\emph{Business horizons}} \bibinfo{volume}{61},
  \bibinfo{number}{4} (\bibinfo{year}{2018}), \bibinfo{pages}{577--586}.
\newblock


\bibitem[\protect\citeauthoryear{Kamar}{Kamar}{2016}]%
        {kamar2016directions}
\bibfield{author}{\bibinfo{person}{Ece Kamar}.}
  \bibinfo{year}{2016}\natexlab{}.
\newblock \showarticletitle{Directions in Hybrid Intelligence: Complementing AI
  Systems with Human Intelligence}. In \bibinfo{booktitle}{\emph{IJCAI}}.
  \bibinfo{pages}{4070--4073}.
\newblock


\bibitem[\protect\citeauthoryear{Klein, Ratliff, Vianello, Adams~Jr,
  Bahn{\'\i}k, Bernstein, Bocian, Brandt, Brooks, Brumbaugh,
  et~al\mbox{.}}{Klein et~al\mbox{.}}{2014}]%
        {klein2014investigating}
\bibfield{author}{\bibinfo{person}{Richard~A Klein}, \bibinfo{person}{Kate~A
  Ratliff}, \bibinfo{person}{Michelangelo Vianello},
  \bibinfo{person}{Reginald~B Adams~Jr},
  \bibinfo{person}{{\v{S}}t{\v{e}}p{\'a}n Bahn{\'\i}k},
  \bibinfo{person}{Michael~J Bernstein}, \bibinfo{person}{Konrad Bocian},
  \bibinfo{person}{Mark~J Brandt}, \bibinfo{person}{Beach Brooks},
  \bibinfo{person}{Claudia~Chloe Brumbaugh}, {et~al\mbox{.}}}
  \bibinfo{year}{2014}\natexlab{}.
\newblock \showarticletitle{Investigating variation in replicability}.
\newblock \bibinfo{journal}{\emph{Social psychology}} (\bibinfo{year}{2014}).
\newblock


\bibitem[\protect\citeauthoryear{Klein, Vianello, Hasselman, Adams, Adams~Jr,
  Alper, Aveyard, Axt, Babalola, Bahn{\'\i}k, et~al\mbox{.}}{Klein
  et~al\mbox{.}}{2018}]%
        {klein2018many}
\bibfield{author}{\bibinfo{person}{Richard~A Klein},
  \bibinfo{person}{Michelangelo Vianello}, \bibinfo{person}{Fred Hasselman},
  \bibinfo{person}{Byron~G Adams}, \bibinfo{person}{Reginald~B Adams~Jr},
  \bibinfo{person}{Sinan Alper}, \bibinfo{person}{Mark Aveyard},
  \bibinfo{person}{Jordan~R Axt}, \bibinfo{person}{Mayowa~T Babalola},
  \bibinfo{person}{{\v{S}}t{\v{e}}p{\'a}n Bahn{\'\i}k}, {et~al\mbox{.}}}
  \bibinfo{year}{2018}\natexlab{}.
\newblock \showarticletitle{Many Labs 2: Investigating variation in
  replicability across samples and settings}.
\newblock \bibinfo{journal}{\emph{Advances in Methods and Practices in
  Psychological Science}} \bibinfo{volume}{1}, \bibinfo{number}{4}
  (\bibinfo{year}{2018}), \bibinfo{pages}{443--490}.
\newblock


\bibitem[\protect\citeauthoryear{Kleinberg, Lakkaraju, Leskovec, Ludwig, and
  Mullainathan}{Kleinberg et~al\mbox{.}}{2018}]%
        {kleinberg2018human}
\bibfield{author}{\bibinfo{person}{Jon Kleinberg}, \bibinfo{person}{Himabindu
  Lakkaraju}, \bibinfo{person}{Jure Leskovec}, \bibinfo{person}{Jens Ludwig},
  {and} \bibinfo{person}{Sendhil Mullainathan}.}
  \bibinfo{year}{2018}\natexlab{}.
\newblock \showarticletitle{Human decisions and machine predictions}.
\newblock \bibinfo{journal}{\emph{The quarterly journal of economics}}
  \bibinfo{volume}{133}, \bibinfo{number}{1} (\bibinfo{year}{2018}),
  \bibinfo{pages}{237--293}.
\newblock


\bibitem[\protect\citeauthoryear{Lai and Tan}{Lai and Tan}{2019}]%
        {lai2019human}
\bibfield{author}{\bibinfo{person}{Vivian Lai} {and} \bibinfo{person}{Chenhao
  Tan}.} \bibinfo{year}{2019}\natexlab{}.
\newblock \showarticletitle{On human predictions with explanations and
  predictions of machine learning models: A case study on deception detection}.
  In \bibinfo{booktitle}{\emph{Proceedings of the conference on fairness,
  accountability, and transparency}}. \bibinfo{pages}{29--38}.
\newblock


\bibitem[\protect\citeauthoryear{Lay and Barbu}{Lay and Barbu}{2010}]%
        {lay2010supervised}
\bibfield{author}{\bibinfo{person}{Nathan Lay} {and} \bibinfo{person}{Adrian
  Barbu}.} \bibinfo{year}{2010}\natexlab{}.
\newblock \showarticletitle{Supervised aggregation of classifiers using
  artificial prediction markets}. In \bibinfo{booktitle}{\emph{ICML}}.
\newblock


\bibitem[\protect\citeauthoryear{Lay and Barbu}{Lay and Barbu}{2012}]%
        {lay2012artificial}
\bibfield{author}{\bibinfo{person}{Nathan Lay} {and} \bibinfo{person}{Adrian
  Barbu}.} \bibinfo{year}{2012}\natexlab{}.
\newblock \showarticletitle{The artificial regression market}.
\newblock \bibinfo{journal}{\emph{arXiv preprint arXiv:1204.4154}}
  (\bibinfo{year}{2012}).
\newblock


\bibitem[\protect\citeauthoryear{Lee, Siewiorek, Smailagic, Bernardino, and
  Berm{\'u}dez~i Badia}{Lee et~al\mbox{.}}{2021}]%
        {lee2021human}
\bibfield{author}{\bibinfo{person}{Min~Hun Lee}, \bibinfo{person}{Daniel~P
  Siewiorek}, \bibinfo{person}{Asim Smailagic}, \bibinfo{person}{Alexandre
  Bernardino}, {and} \bibinfo{person}{Sergi Berm{\'u}dez~i Badia}.}
  \bibinfo{year}{2021}\natexlab{}.
\newblock \showarticletitle{A human-ai collaborative approach for clinical
  decision making on rehabilitation assessment}. In
  \bibinfo{booktitle}{\emph{Proceedings of the 2021 CHI Conference on Human
  Factors in Computing Systems}}. \bibinfo{pages}{1--14}.
\newblock


\bibitem[\protect\citeauthoryear{Lekwijit and Sutivong}{Lekwijit and
  Sutivong}{2018}]%
        {lekwijit2018optimizing}
\bibfield{author}{\bibinfo{person}{Suparerk Lekwijit} {and}
  \bibinfo{person}{Daricha Sutivong}.} \bibinfo{year}{2018}\natexlab{}.
\newblock \showarticletitle{Optimizing the liquidity parameter of logarithmic
  market scoring rules prediction markets}.
\newblock \bibinfo{journal}{\emph{Journal of Modelling in Management}}
  (\bibinfo{year}{2018}).
\newblock


\bibitem[\protect\citeauthoryear{Li, Wang, Zheng, and Franklin}{Li
  et~al\mbox{.}}{2016}]%
        {li2016crowdsourced}
\bibfield{author}{\bibinfo{person}{Guoliang Li}, \bibinfo{person}{Jiannan
  Wang}, \bibinfo{person}{Yudian Zheng}, {and} \bibinfo{person}{Michael~J
  Franklin}.} \bibinfo{year}{2016}\natexlab{}.
\newblock \showarticletitle{Crowdsourced data management: A survey}.
\newblock \bibinfo{journal}{\emph{IEEE Transactions on Knowledge and Data
  Engineering}} \bibinfo{volume}{28}, \bibinfo{number}{9}
  (\bibinfo{year}{2016}), \bibinfo{pages}{2296--2319}.
\newblock


\bibitem[\protect\citeauthoryear{Manski}{Manski}{2006}]%
        {manski2006interpreting}
\bibfield{author}{\bibinfo{person}{Charles~F Manski}.}
  \bibinfo{year}{2006}\natexlab{}.
\newblock \showarticletitle{Interpreting the predictions of prediction
  markets}.
\newblock \bibinfo{journal}{\emph{economics letters}} \bibinfo{volume}{91},
  \bibinfo{number}{3} (\bibinfo{year}{2006}), \bibinfo{pages}{425--429}.
\newblock


\bibitem[\protect\citeauthoryear{Mishina, Murata, Yamauchi, Yamashita, and
  Fujiyoshi}{Mishina et~al\mbox{.}}{2015}]%
        {mishina2015boosted}
\bibfield{author}{\bibinfo{person}{Yohei Mishina}, \bibinfo{person}{Ryuei
  Murata}, \bibinfo{person}{Yuji Yamauchi}, \bibinfo{person}{Takayoshi
  Yamashita}, {and} \bibinfo{person}{Hironobu Fujiyoshi}.}
  \bibinfo{year}{2015}\natexlab{}.
\newblock \showarticletitle{Boosted random forest}.
\newblock \bibinfo{journal}{\emph{IEICE TRANSACTIONS on Information and
  Systems}} \bibinfo{volume}{98}, \bibinfo{number}{9} (\bibinfo{year}{2015}),
  \bibinfo{pages}{1630--1636}.
\newblock


\bibitem[\protect\citeauthoryear{Mohan, Pant, Suyal, and Kumar}{Mohan
  et~al\mbox{.}}{2020}]%
        {mohan2020support}
\bibfield{author}{\bibinfo{person}{Lalit Mohan}, \bibinfo{person}{Janmejay
  Pant}, \bibinfo{person}{Priyanka Suyal}, {and} \bibinfo{person}{Arvind
  Kumar}.} \bibinfo{year}{2020}\natexlab{}.
\newblock \showarticletitle{Support vector machine accuracy improvement with
  classification}. In \bibinfo{booktitle}{\emph{2020 12th International
  Conference on Computational Intelligence and Communication Networks (CICN)}}.
  IEEE, \bibinfo{pages}{477--481}.
\newblock


\bibitem[\protect\citeauthoryear{M{\"u}ller-Schloer and
  Tomforde}{M{\"u}ller-Schloer and Tomforde}{2017}]%
        {muller2017organic}
\bibfield{author}{\bibinfo{person}{Christian M{\"u}ller-Schloer} {and}
  \bibinfo{person}{Sven Tomforde}.} \bibinfo{year}{2017}\natexlab{}.
\newblock \bibinfo{booktitle}{\emph{Organic Computing-Technical Systems for
  Survival in the Real World}}.
\newblock \bibinfo{publisher}{Springer}.
\newblock


\bibitem[\protect\citeauthoryear{Nagar and Malone}{Nagar and Malone}{2011}]%
        {nagar2011making}
\bibfield{author}{\bibinfo{person}{Yiftach Nagar} {and}
  \bibinfo{person}{Thomas~W Malone}.} \bibinfo{year}{2011}\natexlab{}.
\newblock \showarticletitle{Making business predictions by combining human and
  machine intelligence in prediction markets}. Association for Information
  Systems.
\newblock


\bibitem[\protect\citeauthoryear{Nakshatri, Menon, Giles, Rajtmajer, and
  Griffin}{Nakshatri et~al\mbox{.}}{2021}]%
        {nakshatri2021design}
\bibfield{author}{\bibinfo{person}{Nishanth Nakshatri}, \bibinfo{person}{Arjun
  Menon}, \bibinfo{person}{C~Lee Giles}, \bibinfo{person}{Sarah Rajtmajer},
  {and} \bibinfo{person}{Christopher Griffin}.}
  \bibinfo{year}{2021}\natexlab{}.
\newblock \showarticletitle{Design and Analysis of a Synthetic Prediction
  Market using Dynamic Convex Sets}.
\newblock \bibinfo{journal}{\emph{arXiv preprint arXiv:2101.01787}}
  (\bibinfo{year}{2021}).
\newblock


\bibitem[\protect\citeauthoryear{Nunes, Zhang, and Silva}{Nunes
  et~al\mbox{.}}{2015}]%
        {nunes2015survey}
\bibfield{author}{\bibinfo{person}{David~Sousa Nunes}, \bibinfo{person}{Pei
  Zhang}, {and} \bibinfo{person}{Jorge~S{\'a} Silva}.}
  \bibinfo{year}{2015}\natexlab{}.
\newblock \showarticletitle{A survey on human-in-the-loop applications towards
  an internet of all}.
\newblock \bibinfo{journal}{\emph{IEEE Communications Surveys \& Tutorials}}
  \bibinfo{volume}{17}, \bibinfo{number}{2} (\bibinfo{year}{2015}),
  \bibinfo{pages}{944--965}.
\newblock


\bibitem[\protect\citeauthoryear{{Open Science Collaboration}}{{Open Science
  Collaboration}}{2015}]%
        {open2015estimating}
\bibfield{author}{\bibinfo{person}{{Open Science Collaboration}}.}
  \bibinfo{year}{2015}\natexlab{}.
\newblock \showarticletitle{Estimating the reproducibility of psychological
  science}.
\newblock \bibinfo{journal}{\emph{Science}} \bibinfo{volume}{349},
  \bibinfo{number}{6251} (\bibinfo{year}{2015}).
\newblock
\showISSN{0036-8075}
\urldef\tempurl%
\url{https://doi.org/10.1126/science.aac4716}
\showDOI{\tempurl}
\showeprint{https://science.sciencemag.org/content/349/6251/aac4716.full.pdf}


\bibitem[\protect\citeauthoryear{Pawel and Held}{Pawel and Held}{2020}]%
        {pawel2020probabilistic}
\bibfield{author}{\bibinfo{person}{Samuel Pawel} {and}
  \bibinfo{person}{Leonhard Held}.} \bibinfo{year}{2020}\natexlab{}.
\newblock \showarticletitle{Probabilistic forecasting of replication studies}.
\newblock \bibinfo{journal}{\emph{PloS one}} \bibinfo{volume}{15},
  \bibinfo{number}{4} (\bibinfo{year}{2020}), \bibinfo{pages}{e0231416}.
\newblock


\bibitem[\protect\citeauthoryear{Pinto, Kelur, and Shetty}{Pinto
  et~al\mbox{.}}{2018}]%
        {pinto2018iris}
\bibfield{author}{\bibinfo{person}{Joylin~Priya Pinto}, \bibinfo{person}{Soumya
  Kelur}, {and} \bibinfo{person}{Jyothi Shetty}.}
  \bibinfo{year}{2018}\natexlab{}.
\newblock \showarticletitle{Iris flower species identification using machine
  learning approach}. In \bibinfo{booktitle}{\emph{2018 4th International
  Conference for Convergence in Technology (I2CT)}}. IEEE,
  \bibinfo{pages}{1--4}.
\newblock


\bibitem[\protect\citeauthoryear{Polgreen, Nelson, Neumann, and
  Weinstein}{Polgreen et~al\mbox{.}}{2007}]%
        {polgreen2007use}
\bibfield{author}{\bibinfo{person}{Philip~M Polgreen},
  \bibinfo{person}{Forrest~D Nelson}, \bibinfo{person}{George~R Neumann}, {and}
  \bibinfo{person}{Robert~A Weinstein}.} \bibinfo{year}{2007}\natexlab{}.
\newblock \showarticletitle{Use of prediction markets to forecast infectious
  disease activity}.
\newblock \bibinfo{journal}{\emph{Clinical Infectious Diseases}}
  \bibinfo{volume}{44}, \bibinfo{number}{2} (\bibinfo{year}{2007}),
  \bibinfo{pages}{272--279}.
\newblock


\bibitem[\protect\citeauthoryear{Puig, Shu, Li, Wang, Liao, Tenenbaum, Fidler,
  and Torralba}{Puig et~al\mbox{.}}{2020}]%
        {puig2020watch}
\bibfield{author}{\bibinfo{person}{Xavier Puig}, \bibinfo{person}{Tianmin Shu},
  \bibinfo{person}{Shuang Li}, \bibinfo{person}{Zilin Wang},
  \bibinfo{person}{Yuan-Hong Liao}, \bibinfo{person}{Joshua~B Tenenbaum},
  \bibinfo{person}{Sanja Fidler}, {and} \bibinfo{person}{Antonio Torralba}.}
  \bibinfo{year}{2020}\natexlab{}.
\newblock \showarticletitle{Watch-and-help: A challenge for social perception
  and human-ai collaboration}.
\newblock \bibinfo{journal}{\emph{arXiv preprint arXiv:2010.09890}}
  (\bibinfo{year}{2020}).
\newblock


\bibitem[\protect\citeauthoryear{Rajadevi, Devi, Shanthakumari, Latha, Anitha,
  and Devipriya}{Rajadevi et~al\mbox{.}}{2021}]%
        {rajadevi2021feature}
\bibfield{author}{\bibinfo{person}{R Rajadevi}, \bibinfo{person}{EM~Roopa
  Devi}, \bibinfo{person}{R Shanthakumari}, \bibinfo{person}{RS Latha},
  \bibinfo{person}{N Anitha}, {and} \bibinfo{person}{R Devipriya}.}
  \bibinfo{year}{2021}\natexlab{}.
\newblock \showarticletitle{Feature Selection for Predicting Heart Disease
  Using Black Hole Optimization Algorithm and XGBoost Classifier}. In
  \bibinfo{booktitle}{\emph{2021 International Conference on Computer
  Communication and Informatics (ICCCI)}}. IEEE, \bibinfo{pages}{1--7}.
\newblock


\bibitem[\protect\citeauthoryear{Rajpurkar, Chen, Banerjee, and
  Topol}{Rajpurkar et~al\mbox{.}}{2022}]%
        {rajpurkar2022ai}
\bibfield{author}{\bibinfo{person}{Pranav Rajpurkar}, \bibinfo{person}{Emma
  Chen}, \bibinfo{person}{Oishi Banerjee}, {and} \bibinfo{person}{Eric~J
  Topol}.} \bibinfo{year}{2022}\natexlab{}.
\newblock \showarticletitle{AI in health and medicine}.
\newblock \bibinfo{journal}{\emph{Nature Medicine}} \bibinfo{volume}{28},
  \bibinfo{number}{1} (\bibinfo{year}{2022}), \bibinfo{pages}{31--38}.
\newblock


\bibitem[\protect\citeauthoryear{Rajtmajer, Griffin, Wu, Fraleigh, Balaji,
  Squicciarini, Kwasnica, Pennock, McLaughlin, Fritton,
  et~al\mbox{.}}{Rajtmajer et~al\mbox{.}}{2022}]%
        {rajtmajer2022synthetic}
\bibfield{author}{\bibinfo{person}{Sarah Rajtmajer},
  \bibinfo{person}{Christopher Griffin}, \bibinfo{person}{Jian Wu},
  \bibinfo{person}{Robert Fraleigh}, \bibinfo{person}{Laxmaan Balaji},
  \bibinfo{person}{Anna Squicciarini}, \bibinfo{person}{Anthony Kwasnica},
  \bibinfo{person}{David Pennock}, \bibinfo{person}{Michael McLaughlin},
  \bibinfo{person}{Timothy Fritton}, {et~al\mbox{.}}}
  \bibinfo{year}{2022}\natexlab{}.
\newblock \showarticletitle{A synthetic prediction market for estimating
  confidence in published work}. In \bibinfo{booktitle}{\emph{Proceedings of
  the AAAI Conference on Artificial Intelligence}}, Vol.~\bibinfo{volume}{36}.
  \bibinfo{pages}{13218--13220}.
\newblock


\bibitem[\protect\citeauthoryear{Rothschild and Pennock}{Rothschild and
  Pennock}{2014}]%
        {rothschild2014extent}
\bibfield{author}{\bibinfo{person}{David Rothschild} {and}
  \bibinfo{person}{David~M Pennock}.} \bibinfo{year}{2014}\natexlab{}.
\newblock \showarticletitle{The extent of price misalignment in prediction
  markets}.
\newblock \bibinfo{journal}{\emph{Algorithmic Finance}} \bibinfo{volume}{3},
  \bibinfo{number}{1-2} (\bibinfo{year}{2014}), \bibinfo{pages}{3--20}.
\newblock


\bibitem[\protect\citeauthoryear{Singh, Sinha, and Singh}{Singh
  et~al\mbox{.}}{2016}]%
        {singh2016heart}
\bibfield{author}{\bibinfo{person}{Yeshvendra~K Singh}, \bibinfo{person}{Nikhil
  Sinha}, {and} \bibinfo{person}{Sanjay~K Singh}.}
  \bibinfo{year}{2016}\natexlab{}.
\newblock \showarticletitle{Heart disease prediction system using random
  forest}. In \bibinfo{booktitle}{\emph{International Conference on Advances in
  Computing and Data Sciences}}. Springer, \bibinfo{pages}{613--623}.
\newblock


\bibitem[\protect\citeauthoryear{Sowa, Przegalinska, and Ciechanowski}{Sowa
  et~al\mbox{.}}{2021}]%
        {sowa2021cobots}
\bibfield{author}{\bibinfo{person}{Konrad Sowa}, \bibinfo{person}{Aleksandra
  Przegalinska}, {and} \bibinfo{person}{Leon Ciechanowski}.}
  \bibinfo{year}{2021}\natexlab{}.
\newblock \showarticletitle{Cobots in knowledge work: Human--AI collaboration
  in managerial professions}.
\newblock \bibinfo{journal}{\emph{Journal of Business Research}}
  \bibinfo{volume}{125} (\bibinfo{year}{2021}), \bibinfo{pages}{135--142}.
\newblock


\bibitem[\protect\citeauthoryear{Spann and Skiera}{Spann and Skiera}{2009}]%
        {spann2009sports}
\bibfield{author}{\bibinfo{person}{Martin Spann} {and} \bibinfo{person}{Bernd
  Skiera}.} \bibinfo{year}{2009}\natexlab{}.
\newblock \showarticletitle{Sports forecasting: a comparison of the forecast
  accuracy of prediction markets, betting odds and tipsters}.
\newblock \bibinfo{journal}{\emph{Journal of Forecasting}}
  \bibinfo{volume}{28}, \bibinfo{number}{1} (\bibinfo{year}{2009}),
  \bibinfo{pages}{55--72}.
\newblock


\bibitem[\protect\citeauthoryear{Storkey}{Storkey}{2011}]%
        {storkey2011machine}
\bibfield{author}{\bibinfo{person}{Amos Storkey}.}
  \bibinfo{year}{2011}\natexlab{}.
\newblock \showarticletitle{Machine learning markets}. In
  \bibinfo{booktitle}{\emph{Proceedings of the Fourteenth International
  Conference on Artificial Intelligence and Statistics}}.
  \bibinfo{pages}{716--724}.
\newblock


\bibitem[\protect\citeauthoryear{Storkey, Millin, and Geras}{Storkey
  et~al\mbox{.}}{2012}]%
        {storkey2012isoelastic}
\bibfield{author}{\bibinfo{person}{Amos Storkey}, \bibinfo{person}{Jono
  Millin}, {and} \bibinfo{person}{Krzysztof Geras}.}
  \bibinfo{year}{2012}\natexlab{}.
\newblock \showarticletitle{Isoelastic agents and wealth updates in machine
  learning markets}.
\newblock \bibinfo{journal}{\emph{arXiv preprint arXiv:1206.6443}}
  (\bibinfo{year}{2012}).
\newblock


\bibitem[\protect\citeauthoryear{Tetlock}{Tetlock}{2008}]%
        {tetlock2008liquidity}
\bibfield{author}{\bibinfo{person}{Paul~C Tetlock}.}
  \bibinfo{year}{2008}\natexlab{}.
\newblock \showarticletitle{Liquidity and prediction market efficiency}.
\newblock \bibinfo{journal}{\emph{Available at SSRN 929916}}
  (\bibinfo{year}{2008}).
\newblock


\bibitem[\protect\citeauthoryear{Travaini, Pacchioni, Bellumore, Bosia, and
  De~Micco}{Travaini et~al\mbox{.}}{2022}]%
        {travaini2022machine}
\bibfield{author}{\bibinfo{person}{Guido~Vittorio Travaini},
  \bibinfo{person}{Federico Pacchioni}, \bibinfo{person}{Silvia Bellumore},
  \bibinfo{person}{Marta Bosia}, {and} \bibinfo{person}{Francesco De~Micco}.}
  \bibinfo{year}{2022}\natexlab{}.
\newblock \showarticletitle{Machine learning and criminal justice: a systematic
  review of advanced methodology for recidivism risk prediction}.
\newblock \bibinfo{journal}{\emph{International journal of environmental
  research and public health}} \bibinfo{volume}{19}, \bibinfo{number}{17}
  (\bibinfo{year}{2022}), \bibinfo{pages}{10594}.
\newblock


\bibitem[\protect\citeauthoryear{Tschandl, Rinner, Apalla, Argenziano, Codella,
  Halpern, Janda, Lallas, Longo, Malvehy, et~al\mbox{.}}{Tschandl
  et~al\mbox{.}}{2020}]%
        {tschandl2020human}
\bibfield{author}{\bibinfo{person}{Philipp Tschandl},
  \bibinfo{person}{Christoph Rinner}, \bibinfo{person}{Zoe Apalla},
  \bibinfo{person}{Giuseppe Argenziano}, \bibinfo{person}{Noel Codella},
  \bibinfo{person}{Allan Halpern}, \bibinfo{person}{Monika Janda},
  \bibinfo{person}{Aimilios Lallas}, \bibinfo{person}{Caterina Longo},
  \bibinfo{person}{Josep Malvehy}, {et~al\mbox{.}}}
  \bibinfo{year}{2020}\natexlab{}.
\newblock \showarticletitle{Human--computer collaboration for skin cancer
  recognition}.
\newblock \bibinfo{journal}{\emph{Nature Medicine}} \bibinfo{volume}{26},
  \bibinfo{number}{8} (\bibinfo{year}{2020}), \bibinfo{pages}{1229--1234}.
\newblock


\bibitem[\protect\citeauthoryear{Wang, Churchill, Maes, Fan, Shneiderman, Shi,
  and Wang}{Wang et~al\mbox{.}}{2020}]%
        {wang2020human}
\bibfield{author}{\bibinfo{person}{Dakuo Wang}, \bibinfo{person}{Elizabeth
  Churchill}, \bibinfo{person}{Pattie Maes}, \bibinfo{person}{Xiangmin Fan},
  \bibinfo{person}{Ben Shneiderman}, \bibinfo{person}{Yuanchun Shi}, {and}
  \bibinfo{person}{Qianying Wang}.} \bibinfo{year}{2020}\natexlab{}.
\newblock \showarticletitle{From human-human collaboration to Human-AI
  collaboration: Designing AI systems that can work together with people}. In
  \bibinfo{booktitle}{\emph{Extended abstracts of the 2020 CHI conference on
  human factors in computing systems}}. \bibinfo{pages}{1--6}.
\newblock


\bibitem[\protect\citeauthoryear{Wang, Weisz, Muller, Ram, Geyer, Dugan,
  Tausczik, Samulowitz, and Gray}{Wang et~al\mbox{.}}{2019}]%
        {wang2019human}
\bibfield{author}{\bibinfo{person}{Dakuo Wang}, \bibinfo{person}{Justin~D
  Weisz}, \bibinfo{person}{Michael Muller}, \bibinfo{person}{Parikshit Ram},
  \bibinfo{person}{Werner Geyer}, \bibinfo{person}{Casey Dugan},
  \bibinfo{person}{Yla Tausczik}, \bibinfo{person}{Horst Samulowitz}, {and}
  \bibinfo{person}{Alexander Gray}.} \bibinfo{year}{2019}\natexlab{}.
\newblock \showarticletitle{Human-AI collaboration in data science: Exploring
  data scientists' perceptions of automated AI}.
\newblock \bibinfo{journal}{\emph{Proceedings of the ACM on Human-Computer
  Interaction}} \bibinfo{volume}{3}, \bibinfo{number}{CSCW}
  (\bibinfo{year}{2019}), \bibinfo{pages}{1--24}.
\newblock


\bibitem[\protect\citeauthoryear{Wolfers and Zitzewitz}{Wolfers and
  Zitzewitz}{2004}]%
        {wolfers2004prediction}
\bibfield{author}{\bibinfo{person}{Justin Wolfers} {and} \bibinfo{person}{Eric
  Zitzewitz}.} \bibinfo{year}{2004}\natexlab{}.
\newblock \showarticletitle{Prediction markets}.
\newblock \bibinfo{journal}{\emph{Journal of economic perspectives}}
  \bibinfo{volume}{18}, \bibinfo{number}{2} (\bibinfo{year}{2004}),
  \bibinfo{pages}{107--126}.
\newblock


\bibitem[\protect\citeauthoryear{Wolfers and Zitzewitz}{Wolfers and
  Zitzewitz}{2006}]%
        {wolfers2006interpreting}
\bibfield{author}{\bibinfo{person}{Justin Wolfers} {and} \bibinfo{person}{Eric
  Zitzewitz}.} \bibinfo{year}{2006}\natexlab{}.
\newblock \bibinfo{booktitle}{\emph{Interpreting prediction market prices as
  probabilities}}.
\newblock \bibinfo{type}{{T}echnical {R}eport}. \bibinfo{institution}{National
  Bureau of Economic Research}.
\newblock


\bibitem[\protect\citeauthoryear{Wu, Nivargi, Lanka, Menon, Modukuri,
  Nakshatri, Wei, Wang, Caverlee, Rajtmajer, et~al\mbox{.}}{Wu
  et~al\mbox{.}}{2021}]%
        {wu2021predicting}
\bibfield{author}{\bibinfo{person}{Jian Wu}, \bibinfo{person}{Rajal Nivargi},
  \bibinfo{person}{Sree Sai~Teja Lanka}, \bibinfo{person}{Arjun~Manoj Menon},
  \bibinfo{person}{Sai~Ajay Modukuri}, \bibinfo{person}{Nishanth Nakshatri},
  \bibinfo{person}{Xin Wei}, \bibinfo{person}{Zhuoer Wang},
  \bibinfo{person}{James Caverlee}, \bibinfo{person}{Sarah~M Rajtmajer},
  {et~al\mbox{.}}} \bibinfo{year}{2021}\natexlab{}.
\newblock \showarticletitle{Predicting the Reproducibility of Social and
  Behavioral Science Papers Using Supervised Learning Models}.
\newblock \bibinfo{journal}{\emph{arXiv preprint arXiv:2104.04580}}
  (\bibinfo{year}{2021}).
\newblock


\bibitem[\protect\citeauthoryear{Wu, Xiao, Sun, Zhang, Ma, and He}{Wu
  et~al\mbox{.}}{2022}]%
        {wu2022survey}
\bibfield{author}{\bibinfo{person}{Xingjiao Wu}, \bibinfo{person}{Luwei Xiao},
  \bibinfo{person}{Yixuan Sun}, \bibinfo{person}{Junhang Zhang},
  \bibinfo{person}{Tianlong Ma}, {and} \bibinfo{person}{Liang He}.}
  \bibinfo{year}{2022}\natexlab{}.
\newblock \showarticletitle{A survey of human-in-the-loop for machine
  learning}.
\newblock \bibinfo{journal}{\emph{Future Generation Computer Systems}}
  (\bibinfo{year}{2022}).
\newblock


\bibitem[\protect\citeauthoryear{Yang, Youyou, and Uzzi}{Yang
  et~al\mbox{.}}{2020}]%
        {yang2020estimating}
\bibfield{author}{\bibinfo{person}{Yang Yang}, \bibinfo{person}{Wu Youyou},
  {and} \bibinfo{person}{Brian Uzzi}.} \bibinfo{year}{2020}\natexlab{}.
\newblock \showarticletitle{Estimating the deep replicability of scientific
  findings using human and artificial intelligence}.
\newblock \bibinfo{journal}{\emph{Proceedings of the National Academy of
  Sciences}} \bibinfo{volume}{117}, \bibinfo{number}{20}
  (\bibinfo{year}{2020}), \bibinfo{pages}{10762--10768}.
\newblock


\bibitem[\protect\citeauthoryear{Zhu, Wang, and Wang}{Zhu
  et~al\mbox{.}}{2018}]%
        {zhu2018human}
\bibfield{author}{\bibinfo{person}{Meixin Zhu}, \bibinfo{person}{Xuesong Wang},
  {and} \bibinfo{person}{Yinhai Wang}.} \bibinfo{year}{2018}\natexlab{}.
\newblock \showarticletitle{Human-like autonomous car-following model with deep
  reinforcement learning}.
\newblock \bibinfo{journal}{\emph{Transportation research part C: emerging
  technologies}}  \bibinfo{volume}{97} (\bibinfo{year}{2018}),
  \bibinfo{pages}{348--368}.
\newblock


\end{thebibliography}


\end{document}


\pagestyle{fancy}
\fancyhead{}


\maketitle 

\subsection*{(RQ1) Market robustness to hyper-parameters.}

\vspace{0.5cm}
\large \noindent \textbf{Iris classification.} \normalsize

\begin{figure}[h]
  \centering
  \includegraphics[width=0.90\linewidth]{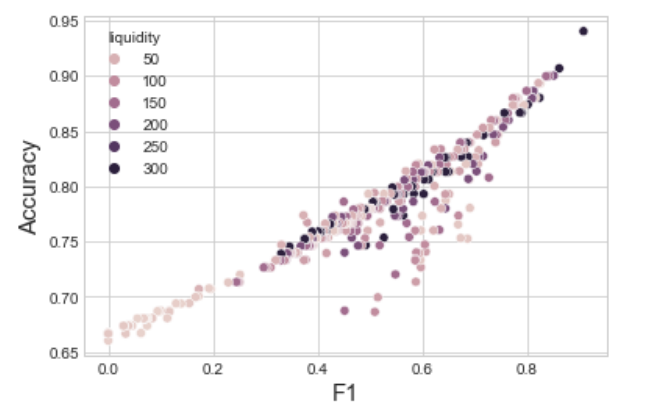}
  \caption{Average F1 score vs. accuracy on the Iris classification task, for varying liquidity.}
  \label{avg_acc_vs_f1_liquidity_iris}
\end{figure}

\begin{figure}[h]
  \centering
  \includegraphics[width=0.90\linewidth]{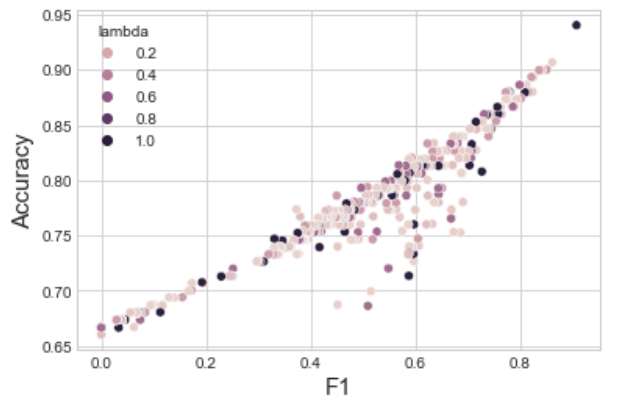}
  \caption{Average F1 score vs. accuracy on the Iris classification task, for varying $\lambda$.}
  \label{avg_acc_vs_f1_lambda_iris}
\end{figure}

\large \noindent \textbf{Heart Disease classification.} \normalsize

\begin{table}[t]
\centering
  \label{table_3_F1_Acc_Lambda}
  \begin{tabular}{ccc|ccc}\toprule
    \textbf{$\lambda$} & Liquidity & Cash & Accuracy & F1 & Scored \% \\ 
    \midrule
    \textbf{0.01} & 50 & 20 & 0.55 & 0.64 & 99.67\\
    \textbf{0.025} & 50 & 20 & 0.56 & 0.64 & 99.67\\
    \textbf{0.05} & 50 & 20 & 0.66 & 0.71 & 99.67\\
    \textbf{0.1} & 50 & 20 & 0.60 & 0.66 & 100\\
    \textbf{0.25} & 50 & 20 & 0.58 & 0.67 & 99.67\\
    \textbf{0.5} & 50 & 20 & 0.60 & 0.65 & 99.34\\ 
    \textbf{1.0} & 50 & 20 & 0.58 & 0.63 & 99.67\\
    \bottomrule
  \end{tabular}
    \caption{Average F1 score and accuracy on the Heart Disease classification task, varying Lambda.}
\end{table}

\begin{table}[t]
\centering
  \label{table_3_F1_Acc_Initial Cash}
  \begin{tabular}{ccc|ccc}\toprule
    \textbf{Cash} & Liquidity & $\lambda$ & Accuracy & F1 & Scored \% \\ 
    \midrule
    \textbf{1} & 50 & 0.05 & 0.47 & 0.38 & 100\\
    \textbf{2} & 50 & 0.05 & 0.55 & 0.56 & 100\\
    \textbf{3} & 50 & 0.05 & 0.58 & 0.64 & 100\\
    \textbf{4} & 50 & 0.05 & 0.59 & 0.64 & 100\\
    \textbf{5} & 50 & 0.05 & 0.58 & 0.61 & 99.01\\
    \textbf{10} & 50 & 0.05 & 0.60 & 0.63 & 99.34\\ 
    \textbf{20} & 50 & 0.05 & 0.66 & 0.71 & 99.67\\
    \bottomrule
  \end{tabular}
    \caption{Average F1 score and accuracy on the Heart Disease classification task, varying Initial Cash.}
\end{table}

\begin{figure}[h]
  \centering
  \includegraphics[width=0.90\linewidth]{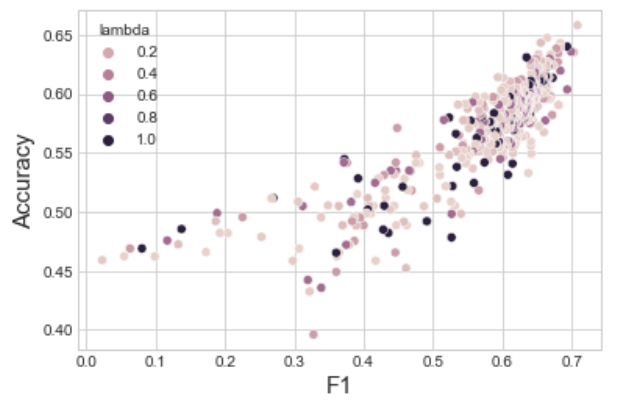}
  \caption{Average F1 score vs. accuracy on the Heart Disease classification task, for varying $\lambda$.}
  \label{avg_acc_vs_f1_lambda_heart}
\end{figure}

\begin{figure}[h]
  \centering
  \includegraphics[width=0.90\linewidth]{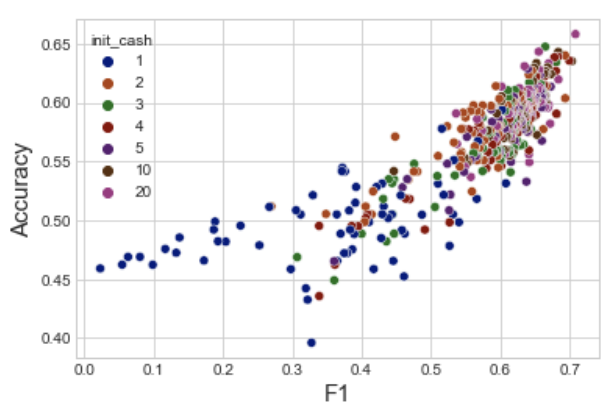}
  \caption{Average F1 score vs. accuracy on the Heart Disease classification task, for varying Initial Cash.}
  \label{avg_acc_vs_f1_lambda_heart}
\end{figure}

\vspace{0.5cm}
\large \noindent \textbf{Replication prediction.} \normalsize

\begin{figure}[h]
  \centering
  \includegraphics[width=0.90\linewidth]{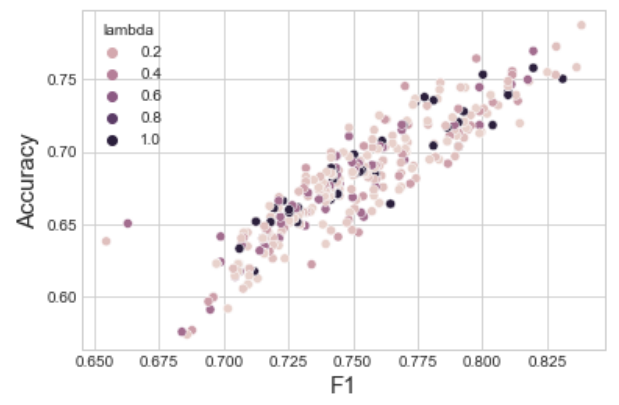}
  \caption{Average F1 score vs. accuracy with different values of $\lambda$ for Replication Data}
    \label{avg_acc_vs_f1_liquidity_replication}
\end{figure}

\begin{figure}[h]
  \centering
  \includegraphics[width=0.90\linewidth]{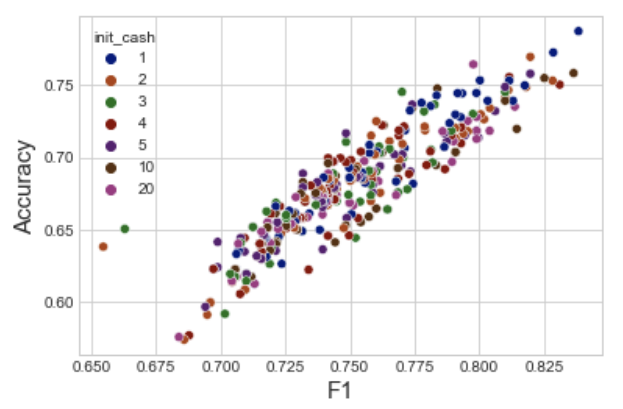}
  \caption{Average F1 score vs. accuracy with different values of Initial cash for Replication Data}
    \label{avg_acc_vs_f1_liquidity_replication}
\end{figure}

\begin{table}[t]
\centering
  \caption{Average F1 score and accuracy on the Replication classification task, varying Initial Cash.}
  \label{table_3_F1_Acc_liquidity_heart}
  \begin{tabular}{ccc|ccc}\toprule
    \textbf{Cash} & Liquidity & $\lambda$ & Accuracy & F1 & Scored \% \\ 
    \midrule
    \textbf{1} & 5 & 0.05 & 0.79 & 0.84 & 35.86\\
    \textbf{2} & 5 & 0.05 & 0.65 & 0.75 & 35.86\\
    \textbf{3} & 5 & 0.05 & 0.67 & 0.76 & 35.86\\
    \textbf{4} & 5 & 0.05 & 0.66 & 0.74 & 37.24\\
    \textbf{5} & 5 & 0.05 & 0.70 & 0.78 & 36.55\\
    \textbf{10} & 5 & 0.05 & 0.66 & 0.76 & 36.55\\ 
    \textbf{20} & 5 & 0.05 & 0.67 & 0.75 & 36.55\\
    \bottomrule
  \end{tabular}
\end{table}

\begin{table}[t]
\centering
  \caption{Average F1 score and accuracy on the Replication classification task, varying Lambda.}
  \label{table_3_F1_Acc_liquidity_heart}
  \begin{tabular}{ccc|ccc}\toprule
    \textbf{$\lambda$} & Liquidity & Cash & Accuracy & F1 & Scored \% \\ 
    \midrule
    \textbf{0.01} & 5 & 1 & 0.74 & 0.80 & 36.55\\
    \textbf{0.025} & 5 & 1 & 0.76 & 0.82 & 35.86\\
    \textbf{0.05} & 5 & 1 & 0.79 & 0.84 & 35.86\\
    \textbf{0.1} & 5 & 1 & 0.77 & 0.83 & 35.86\\
    \textbf{0.25} & 5 & 1 & 0.73 & 0.79 & 37.24\\
    \textbf{0.5} & 5 & 1 & 0.75 & 0.80 & 36.55\\ 
    \textbf{1.0} & 5 & 1 & 0.73 & 0.79 & 37.24\\
    \bottomrule
  \end{tabular}
\end{table}

\begin{table}[t]
\centering
  \caption{Average F1 score and accuracy on the Replication classification task, varying Liquidity.}
  \label{table_3_F1_Acc_liquidity}
  \begin{tabular}{ccc|ccc}\toprule
    \textbf{Liquidity} & $\lambda$ & Cash & Accuracy & F1 & Scored \% \\ 
    \midrule
    \textbf{5} & 0.05 & 1 & 0.79 & 0.84 & 35.86\\
    \textbf{10} & 0.05 & 1 & 0.74 & 0.81 & 36.55\\
    \textbf{20} & 0.05 & 1 & 0.62 & 0.70 & 36.55\\
    \textbf{50} & 0.05 & 1 & 0.64 & 0.72 & 37.24\\
    \textbf{75} & 0.05 & 1 & 0.65 & 0.72 & 35.17\\
    \textbf{100} & 0.05 & 1 & 0.71 & 0.76 & 35.17\\ 
    \textbf{150} & 0.05 & 1 & 0.71 & 0.77 & 39.31\\
    \textbf{200} & 0.05 & 1 & 0.69 & 0.75 & 39.31\\
    \textbf{300} & 0.05 & 1 & 0.74 & 0.78 & 39.31\\
    \bottomrule
  \end{tabular}
\end{table}

\begin{figure}[!h]
  \centering
  \includegraphics[scale=0.4]{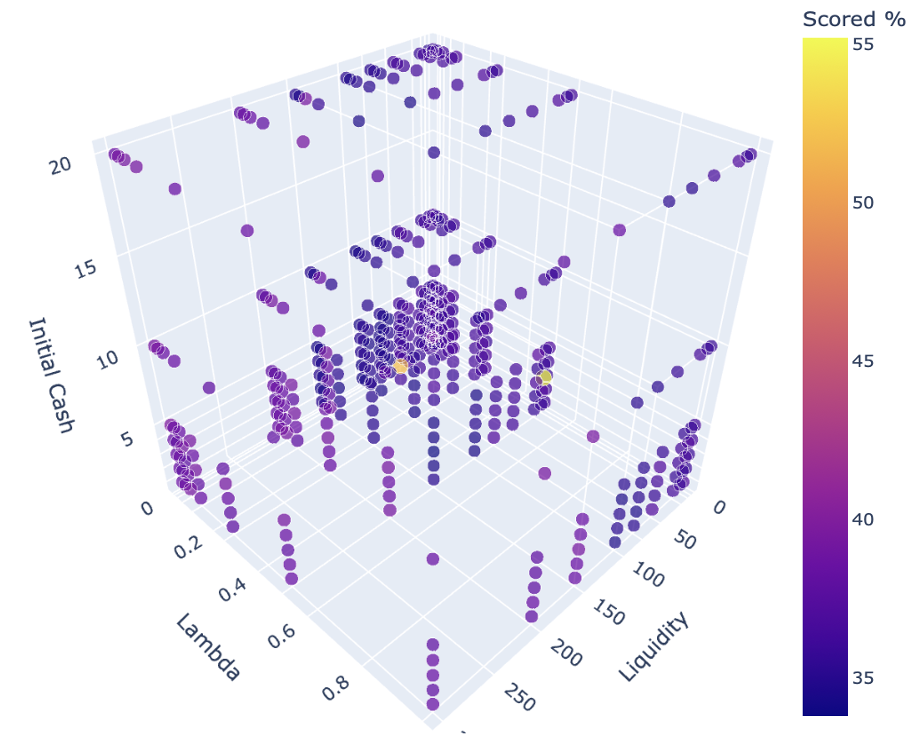}
  \caption{Plot of liquidity, lambda, Initial Cash vs Scored \% for replication data}
  \label{4d_replication_scored_1}
\end{figure}

\vspace{1cm}
\subsection*{(RQ2) Market behavior with exogenous agents.}
\vspace{1cm}

\begin{figure}[h]
  \centering
  \includegraphics[width=0.95\linewidth]{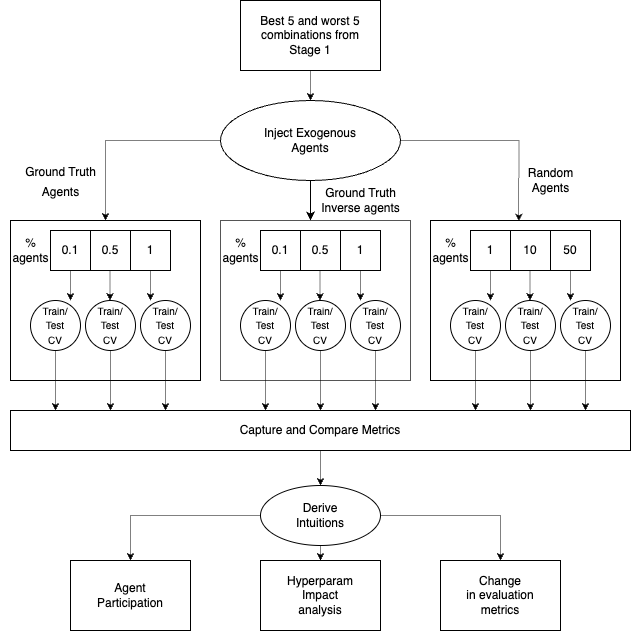}
  \caption{RQ2 experimental architecture.}
  \label{phase_2_arch_su}
\end{figure}

\begin{table*}[t]
  \caption{Average accuracy scores on 10 replication prediction markets, for different types and size of exogenous agent populations}
  \label{table_f1_exogenous}
  \begin{tabular}{c|ccc|ccc|cccc}\toprule
    None & GT 0.1\% & GT 0.5\% & GT 1\% &  GTinv 0.1\% & GTinv 0.5\% & GTinv 1\% &  Random 1\% & Random 5\% & Random 10\% & Random 50\%\\ \midrule
    0.79 & 0.92 & 1 & 1 & 0.27 & 0.14 & 0.05 & 0.76 & 0.77 & 0.77 & 0.78 \\
    0.76 & 0.90 & 0.95 & 0.97 & 0.27 & 0.21 & 0.17 & 0.67 & 0.68 & 0.67 & 0.70 \\
    0.75 & 0.94 & 1 & 1 & 0.26 & 0.14 & 0.03 & 0.76 & 0.77 & 0.80 & 0.77 \\
    0.77 & 0.89 & 0.95 & 0.99 & 0.26 & 0.18 & 0.14 & 0.70 & 0.66 & 0.77 & 0.73 \\
    0.75 & 0.87 & 0.90 & 0.94 & 0.27 & 0.23 & 0.18 & 0.67 & 0.70 & 0.74 & 0.68 \\ \midrule
    0.58 & 0.88 & 0.93 & 0.95 & 0.26 & 0.21 & 0.17 & 0.70 & 0.70 & 0.76 & 0.71 \\ 
    0.57 & 0.90 & 0.96 & 0.95 & 0.25 & 0.18 & 0.18 & 0.73 & 0.70 & 0.71 & 0.72 \\ 
    0.58 & 0.89 & 0.94 & 0.95 & 0.25 & 0.19 & 0.17 & 0.73 & 0.71 & 0.73 & 0.74 \\ 
    0.65 & 0.88 & 0.89 & 0.93 & 0.25 & 0.26 & 0.18 & 0.71 & 0.68 & 0.74 & 0.69 \\ 
    0.64 & 0.89 & 0.90 & 0.94 & 0.27 & 0.23 & 0.19 & 0.72 & 0.68 & 0.76 & 0.70 \\ \bottomrule
  \end{tabular}
\end{table*}

\begin{figure}[h]
  \centering
  \includegraphics[width=0.90\linewidth]{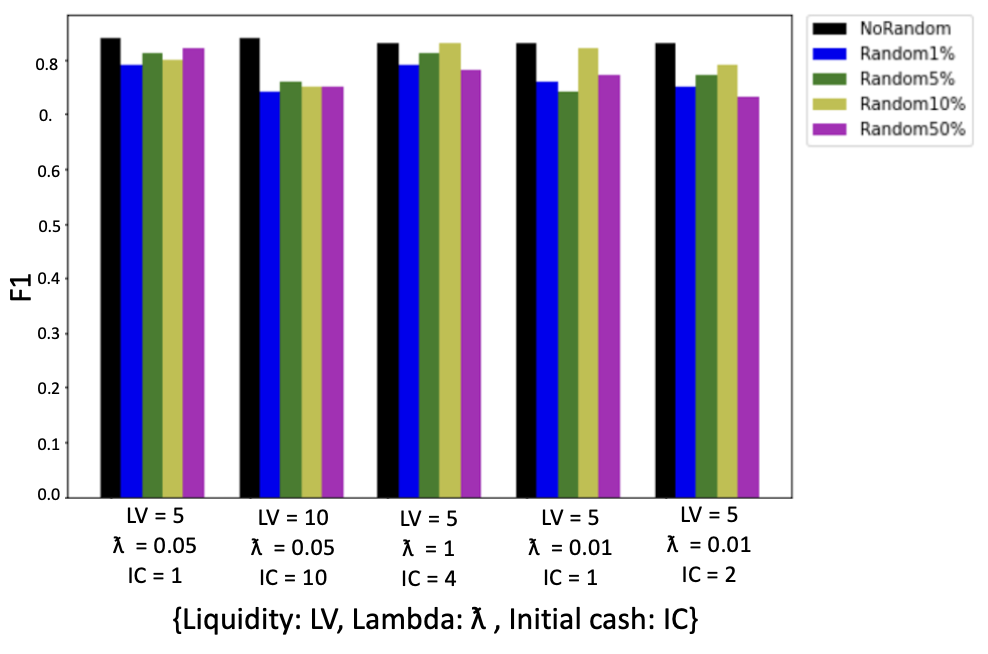}
  \caption{F1 with different percentage of added Random agents for replication data}
  \label{f1_random_agents}
\end{figure}

\begin{figure}[!h]
  \centering
  \includegraphics[width=0.90\linewidth]{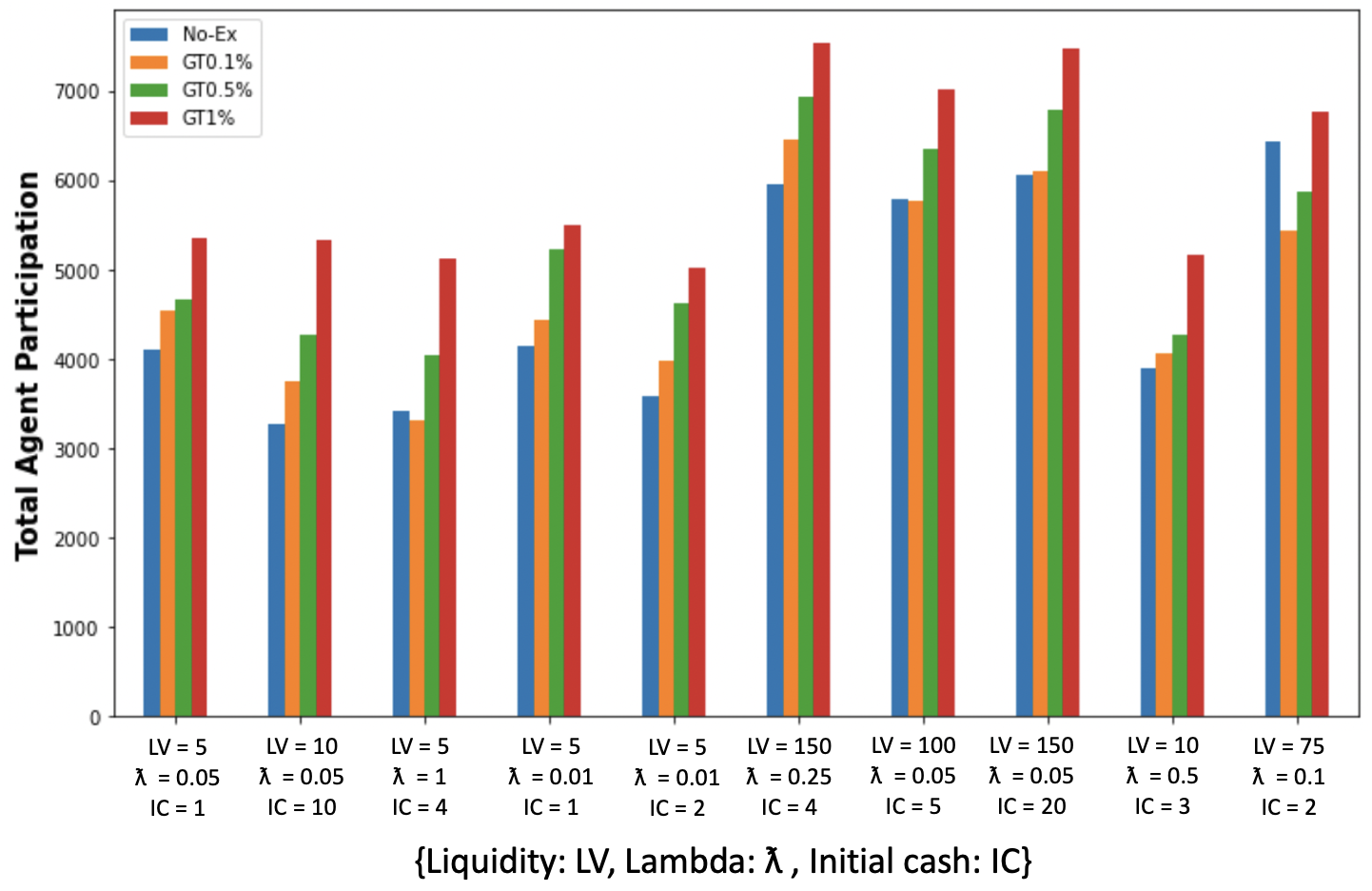}
  \caption{Agent Participation with different percentage of added GT agents for replication data}
   \label{participation_gt}
\end{figure}

\begin{figure}[!h]
  \centering
  \includegraphics[width=0.90\linewidth]{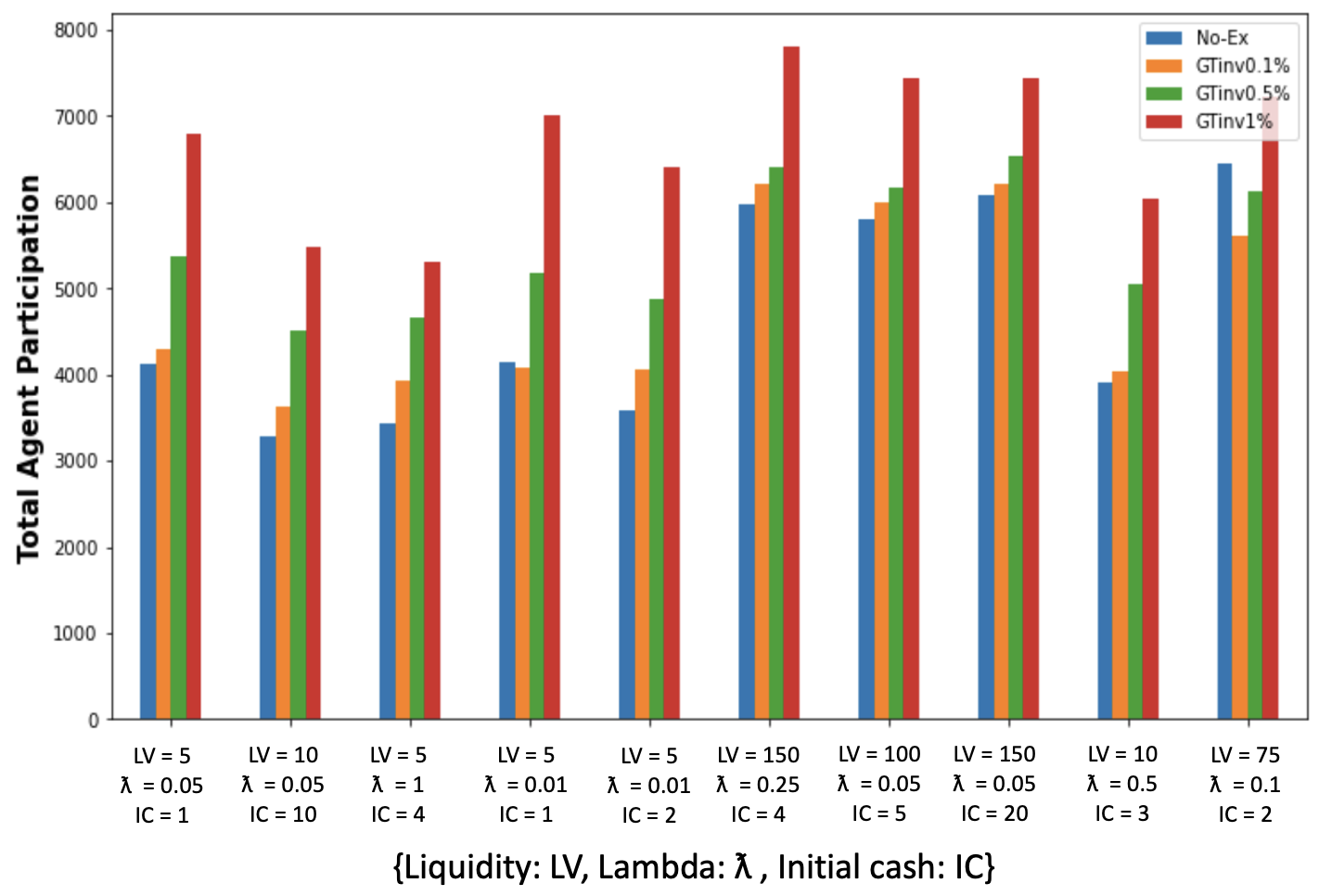}
  \caption{Agent Participation with different percentage of added GT inverse agents for replication data}
  \label{participation_gt_inv}
\end{figure}